\documentclass{cup-hpl}

\usepackage{comment}
\usepackage[table]{xcolor}
\usepackage{booktabs}
\usepackage{array}
\usepackage{makecell}
\usepackage{url}

\newcommand{\equalcontrib}{\ensuremath{^{\dagger}}}

\begin{document}

\shorttitle{ELIMAIA--ELIMED dosimetric commissioning}
\shortauthor{G.A.P. Cirrone et al.}

\title{Relative and absolute dosimetric commissioning of the ELIMAIA--ELIMED laser-driven ion beamline for 23.5 MeV proton beams}

\author[1,2,3]{G.A.P. Cirrone\equalcontrib{}\corresp{Dr G.A. Pablo Cirrone (pablo.cirrone@lns.infn.it) and Dr Giada Petringa (giada.petringa@lns.infn.it). \ensuremath{^{\dagger}}G.A.P. Cirrone and G. Petringa contributed equally to this work and should be considered co-first authors.}}

\author[1]{G. Petringa\equalcontrib{}}
\author[1]{A. Kurmanova}
\author[1]{R. Catalano}
\author[1]{A. Amato}
\author[1,4]{A. Pappalardo}
\author[1]{M.G. Guarrera}
\author[5]{F. Schillaci}
\author[5]{P. Blaha}
\author[5]{F. Grepl}
\author[5]{M. Tryus}
\author[5]{A. Velyhan}
\author[5]{L. Giuffrida}
\author[5]{D. Margarone}
\author[1]{G. Cuttone}

\address[1]{Istituto Nazionale di Fisica Nucleare - Laboratori Nazionali del Sud - Via S. Sofia 62, Catania, Italy}
\address[2]{Centro Siciliano di Fisica Nucleare e Struttura della Materia - Via S. Sofia 64, Catania, Italy}
\address[3]{The Extreme Light Infrastructure ERIC, Za Radnicí 835, 252 41 Dolní Brežany, Czech Republic}
\address[4]{Extreme Light Infrastructure - Nuclear Physics (ELI-NP), Horia Hulubei National Institute for R\&D in Physics and Nuclear Engineering (IFIN-HH), 30 Reactorului Street, 077125 Măgurele, Romania}
\address[5]{ELI Beamlines Facility - The Extreme Light Infrastructure ERIC, Za Radnicí 835, 252 41 Dolní Brežany, Czech Republic}

\begin{abstract}
The development of laser-driven proton beamlines for biomedical and radiobiological applications requires traceable dosimetry and reliable online monitoring at the irradiation point. In this work, we report the relative and absolute dosimetric commissioning of the ELIMAIA--ELIMED laser-driven proton beamline at ELI Beamlines using an energy-selected proton beam with an average energy of about $23.5~\mathrm{MeV}$. Radiochromic-film measurements were used to characterize the transverse dose distribution, the depth--dose profile, and the proton energy spectrum at the irradiation point. The reconstructed spectrum was centred at $23.5~\mathrm{MeV}$ with a FWHM of $2.60~\mathrm{MeV}$, while the transverse dose distribution showed an approximately $5.5~\mathrm{mm}$ field size with a millimetre-scale homogeneous region. A complementary diamond time-of-flight measurement provided an independent check of the spectral reconstruction and gave a proton-bunch temporal FWHM of $1.70~\mathrm{ns}$, corresponding to an estimated peak dose rate of about $4.0\times10^{6}~\mathrm{Gy\,s^{-1}}$ under the present irradiation conditions. The Faraday Cup was used as the absolute reference detector for dose-to-water determination and for cross-calibrating the Dual-Gap Ionization Chamber, which was operated as the primary online dose monitor. The Integrating Current Transformer and Secondary Electron Monitor were evaluated as upstream relative beam monitors. An independent RCF--FC dose comparison over fifty consecutive shots yielded $30.45 \pm 3.50~\mathrm{cGy}$ from the FC and $36.46 \pm 1.80~\mathrm{cGy}$ from the EBT3 film. Dedicated G4ELIMED Monte Carlo simulations showed that the RCF positioned upstream of the FC perturbs the FC measurement through proton losses outside the FC acceptance and beam broadening at the FC entrance. Applying the resulting correction factor, $C_{\mathrm{MC}}=1.109$, reduced the residual difference between the FC- and RCF-derived doses to about $7\%$. These results establish the dosimetric chain of the ELIMAIA--ELIMED beamline under the present low-fluence commissioning conditions and identify the operational limits of the online monitoring system.
\end{abstract}

\maketitle

\tableofcontents
\newpage


\section{Introduction}
The interaction of intense laser pulses with solid targets can generate multi-MeV proton and ion beams with ultrashort duration, high brightness, and high particle density. Depending on the laser and target parameters, ion acceleration may occur through different mechanisms, including Target Normal Sheath Acceleration (TNSA), radiation-pressure acceleration, collisionless-shock acceleration, and relativistic-transparency regimes, each of which produces distinct spectral, angular, and temporal beam characteristics~\cite{Macchi2013}. These properties make laser-driven ion sources attractive for a broad range of multidisciplinary applications, including detector development, material irradiation, radiation-effects studies, dosimetry, and radiobiology. In particular, laser-driven proton beams provide access to temporal dose-delivery regimes that are difficult to reproduce with conventional accelerators, making them relevant for investigations of ultra-high-dose-rate radiation effects~\cite{Chaudhary2021,Hideghty2025}.

The generation of energetic particles at the laser target, however, represents only the first step toward their practical use. Laser-driven proton beams are generally characterized by broad energy spectra, large angular divergence, and shot-to-shot variations in particle number, spectral distribution, and pointing. Consequently, the source beam cannot normally be delivered directly to an experimental sample. Dedicated systems are required to collect and transport the particles, select the desired energy interval, shape the spatial distribution, and monitor the delivered beam on a shot-by-shot basis~\cite{Petringa2020,Milluzzo2018,Schillaci2022,brack2020spectral,ziegler2021proton}. These requirements are particularly stringent for radiobiological experiments, where interpretation of the biological endpoint requires accurate knowledge of the absorbed dose and of the spatial, spectral, and temporal properties of the irradiation field. Reliable radiobiology therefore requires a complete irradiation platform that integrates controlled beam transport, reproducible field formation, absolute and relative dosimetry, and online beam monitoring.

Several research groups have made important progress toward this objective. At the Berkeley Lab Laser Accelerator facility, compact focusing systems have been used to transport laser-driven protons to an irradiation station and to perform radiobiological experiments at high repetition rate, demonstrating the feasibility of controlled beam delivery from a laser-driven source~\cite{bin2022new}. A systematic development in this direction has also been carried out at the Dresden platform, where the DRACO laser-driven proton source has been progressively integrated with dedicated transport, spectral-selection, field-shaping, monitoring, and dosimetric systems~\cite{brack2020spectral,ziegler2021proton,metzkes2023}. This platform enabled the delivery of spatially homogenized and quantitatively controlled proton fields to millimetre-scale biological targets, and was subsequently used for dose-controlled tumour irradiation in mice and for further \textit{in vivo} investigations under ultra-high-dose-rate conditions~\cite{Kroll2022,metzkes2023}. A dedicated dosimetric framework was also established by combining independent dose measurements with transmission monitoring of pulse-to-pulse spectral variations, allowing the corresponding changes in the depth--dose distribution to be evaluated during irradiation~\cite{Reimold2023}. These achievements demonstrate that meaningful radiobiological studies become possible only when the laser-driven beam has reached an adequate level of stability, characterization, and dosimetric control.

Within this context, the Extreme Light Infrastructure has identified the development of reliable irradiation platforms as a central element of its roadmap for radiobiology and cancer research. The objective is to exploit the distinctive temporal and spatial characteristics of laser-generated radiation while providing users with controlled, reproducible, and traceable irradiation conditions suitable for systematic biological investigations~\cite{Hideghty2025}.

At ELI Beamlines, the ELIMAIA (ELI Multidisciplinary Applications of laser--Ion Acceleration)--ELIMED (ELI MEDical Applications) beamline was specifically conceived to support this transition from laser-driven particle generation to application-ready ion beams~\cite{margarone2018elimaia,cirrone2020elimed,schillaci2022elimaia,giuffrida2025elimaia}. A distinctive feature of ELIMAIA--ELIMED is that it is operated as a user-oriented beamline within the ELI research infrastructure. Experimental access is provided through peer-reviewed user programmes~\cite{ELIUserPortal2026}, allowing the broader medical-physics, radiobiology, and multidisciplinary communities to use a dedicated laser-driven ion irradiation platform rather than a system developed only for project-specific experiments or internal institutional use.

The ELIMAIA section comprises the laser--target interaction, target-handling, and source-diagnostic systems, whereas the ELIMED section is designed to collect, transport, energy-select, shape, and monitor the accelerated ions before their delivery to an in-air user station. The magnetic transport system, together with the dedicated diagnostic and dosimetric instrumentation, is intended to provide beams with controlled energy, fluence, and spatial distribution. Establishing and validating this capability is an essential prerequisite for multidisciplinary applications and, in particular, for reproducible radiobiological irradiations.

The platform architecture was conceived for progressive scaling toward operation at higher repetition rate, matching the nominal capabilities of the L3-HAPLS driver. Although the present commissioning was performed under lower-fluence and lower-repetition-rate conditions, the integrated design of ELIMAIA--ELIMED is intended to support controlled sequences of proton bunches in future campaigns. Such operation would enable multi-pulse or fast-fractionation irradiation schemes and relatively high mean dose rates, while retaining the extremely high instantaneous dose rate characteristic of individual laser-driven proton bunches.

In this work, we report the first relative and absolute dosimetric commissioning of the ELIMAIA--ELIMED beamline using an energy-selected laser-driven proton beam with an average energy of approximately $24~\mathrm{MeV}$ transported to the in-air irradiation point. The beam was characterized in terms of transverse dose distribution, depth--dose profile, energy spectrum, fluence, and absorbed dose to water. A complementary detector chain, comprising a Faraday Cup (FC), a Dual-Gap Ionization Chamber (DGIC), an Integrating Current Transformer (ICT), a Secondary Electron Monitor (SEM), and calibrated radiochromic films, was used to establish an absolute dose reference and to cross-calibrate the online monitoring systems. This commissioning represents a necessary step toward the delivery of stable, quantitatively characterized, and reproducible proton beams for future radiobiological user experiments at ELIMAIA--ELIMED.

The remainder of this paper is organized as follows. Section~\ref{elimed-elimaia} describes the ELIMAIA--ELIMED facility, including the laser-driven proton source, the magnetic transport system, and the dosimetric section of the beamline. Section~\ref{materials and methods} presents the experimental configuration, the detector systems, and the procedures adopted for signal processing and dose determination. Section~\ref{sec:results} reports the relative and absolute dosimetric characterization of the transported proton beam, including its spatial and spectral properties and the cross-calibration of the online beam monitors. Finally, Section~\ref{conclusion} summarizes the main results and discusses the perspectives for future radiobiological experiments at ELIMAIA--ELIMED.

\section{ELIMAIA--ELIMED beamline}\label{elimed-elimaia}

\subsection{Proton acceleration at ELIMAIA}\label{proton acceleration}

The dosimetric commissioning reported in this study was performed at the ELIMAIA--ELIMED beamline in the E4 Ion Acceleration Hall of ELI Beamlines~\cite{margarone2018elimaia,cirrone2020elimed}.

During the campaign, proton bunches were produced by the L3-HAPLS PW-class laser delivering $\sim 10~\mathrm{J}$, $27~\mathrm{fs}$ pulses focused onto $6~\mu\mathrm{m}$-thick metallic foils. The resulting on-target peak intensity was approximately $3\times10^{21}~\mathrm{W\,cm^{-2}}$, consistent with TNSA-dominated acceleration. This operating point lies within the broader L3-HAPLS capabilities described in the introduction, namely pulse energies up to $\approx 30~\mathrm{J}$, pulse durations below $30~\mathrm{fs}$, and repetition rates up to $10~\mathrm{Hz}$~\cite{margarone2018elimaia,cirrone2020elimed}.

\begin{figure*}
    \centering
    \includegraphics[width=\linewidth]{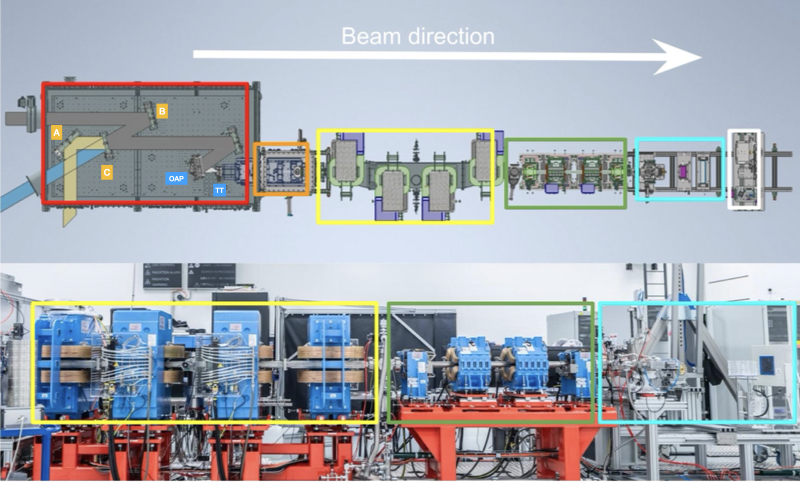}
    \caption{(a) Schematic layout of the ELIMAIA--ELIMED beamline. 1: interaction chamber (red); 2: collecting and focusing section (orange); 3: energy selection section (yellow); 4: conventional transport section (green); 5: in-air section (cyan); 6: irradiation point (white). (b) Photograph of the corresponding beamline elements, highlighted with the same colour code as in panel (a). The beam propagates from left to right, as indicated by the arrow.}
    \label{fig:top_view_elimaia}
\end{figure*}

Figure~\ref{fig:top_view_elimaia} shows a simplified top view of the interaction chamber together with the ELIMED transport and irradiation sections. Three flat mirrors (labelled A, B, and C) steer and centre the laser beam onto an $f/1.75$ off-axis parabola (OAP), yielding a near diffraction-limited focal spot of approximately $1.5 \times 1.5~\mu\mathrm{m}^2$ (FWHM) at the interaction point. Target positioning and movement are provided by a motorized Target Tower (TT), which accommodates up to nine frames, each hosting up to 100 targets; the tower can be remotely controlled and automatically indexed shot by shot by the laser control system~\cite{giuffrida2025elimaia}.

Protons generated at the TT are injected into the ELIMED line. They are first captured and refocused in the collection/focusing section to reduce their initial divergence and to match the acceptance of the downstream optics. They are then energy-selected in the Energy Selection System (ESS) chicane. After transport, the selected beam is extracted into air through the Kapton window and propagates through the in-air dosimetry section, which hosts transport elements and online diagnostics, up to the user irradiation point, as detailed in the following subsections.

\subsection{ELIMED beam transport and dosimetry section}\label{elimed}

Laser-driven ion beams are broadband and highly divergent; the ELIMED section is therefore organized into three consecutive blocks that (i) collect and refocus the particles, (ii) select the desired energy range, and (iii) condition and extract the beam toward the in-air dosimetry and irradiation station. A view of the full line is given in Figure~\ref{fig:top_view_elimaia}.
Immediately downstream of the interaction chamber, ions are captured in a dedicated vacuum vessel hosting five Permanent Magnet Quadrupoles (PMQs). This section reduces the initial divergence in both transverse planes and matches the beam to the acceptance of the energy selector. The PMQs, which provide gradients up to $\sim100~\mathrm{T/m}$ over a $36~\mathrm{mm}$ bore, are mounted on a displacement system that allows the optics to be retuned for different ion species and energies~\cite{Schillaci2020}. The optics form an image at the dispersive plane where energy selection is performed, while simultaneously reducing the angular aperture for optimal transmission through the selector.
Energy selection is performed by a double-dispersive magnetic chicane composed of four C-shaped resistive dipoles with laminated cores operated in alternating polarity. A rectangular slit placed after the second dipole trims the energy spread around the set point~\cite{schillaci2016design, Schillaci2020}. The dipole fields are adjustable to accommodate the chosen ion species and energy; in nominal configuration, the ESS can select protons up to 300~AMeV and heavier ions up to 60~AMeV/u.
Downstream of the selector, two electromagnetic quadrupoles and two steerers provide the final spot shaping and centring of the selected beam (``conventional transport section'' in Figure~\ref{fig:top_view_elimaia}). The beam is then extracted into air through a $50~\mu\mathrm{m}$ Kapton window toward the dosimetry station and user plane.
Along the short in-air path, a complementary set of diagnostics measures and monitors the beam on a shot-by-shot basis up to the irradiation point: an Integrating Current Transformer (ICT) positioned in vacuum at the window, a Secondary Electron Monitor (SEM) located a few centimetres downstream in air, and a Dual-Gap Ionization Chamber (DGIC) acting as a calibrated dose monitor with recombination correction. A Faraday Cup (FC) placed at the irradiation point provides the absolute reference, while stacks of calibrated radiochromic films (RCFs) are used to assess transverse uniformity and to reconstruct depth--dose profiles and energy spectra when needed. Time-of-flight (TOF) diagnostics, based on diamond and silicon-carbide detectors, are also employed along the line for spectral and timing information~\cite{scuderi2017time,milluzzo2019new}.
With this layout, ELIMED is designed to deliver stable, reproducible, and tunable ion beams at the user location, both in energy and fluence, while providing the instrumentation and procedures required for robust absolute and relative dosimetry and for radiobiological irradiations.

\section{Experimental configuration and measurement approach}\label{materials and methods}

\subsection{Beamline configuration}\label{beamline config}

Protons accelerated in the interaction chamber were first transported through the in-vacuum section. The beam optics, including the PMQs, the ESS, and the final steerers, were tuned to deliver proton bunches with an energy distribution centred at about $24~\mathrm{MeV}$. The operational settings of the electromagnetic transport elements and their positions relative to the laser interaction point are reported in Table~\ref{tab:tab1}.

\begin{table*}[ht]
\centering
\caption{Main electromagnetic and geometrical characteristics of the beam transport elements as they were set during the measurements presented in this paper.}
\label{tab:tab1}
\begin{tabular}{|p{3cm}|p{3cm}|p{3cm}|p{3cm}|}
\hline
\rowcolor{gray!20} 
\multicolumn{4}{|c|}{Collecting and focusing section} \\
\hline
\centering Element & \centering Length [mm] & \centering Magnetic field gradient [T/m] & \parbox[c]{3cm}{\centering Distance from the source [mm]} \\

\hline
Permanent magnetic quadrupole 1 & \centering 120 &  \parbox[c]{3cm}{\centering 99} & \parbox[c]{3cm}{\centering 79} \\
\hline
Permanent magnetic quadrupole 2 & \parbox[c]{3cm} {\centering 80} & \parbox[c]{3cm} {\centering -94} & \parbox[c]{3cm} {\centering 142} \\
\hline
Permanent magnetic quadrupole 3 & \parbox[c]{3cm} {\centering 80} & \parbox[c]{3cm} {\centering 94} & \parbox[c]{3cm} {\centering 277.2} \\
\hline
Permanent magnetic quadrupole 4 & \parbox[c]{3cm} {\centering 120} & \parbox[c]{3cm} {\centering -99} & \parbox[c]{3cm} {\centering 353.7} \\
\hline
\rowcolor{gray!20} 
\multicolumn{4}{|c|}{Energy selection section} \\
\hline
& Dipole magnetic field [T] & Slit aperture [mm] &\\
\hline
All four dipoles & \parbox[c]{3cm} {\centering 0.26}& \parbox[c]{3cm} {\centering 30} & \parbox[c]{3cm} {\centering --}\\ 
\hline
\rowcolor{gray!20} 
\multicolumn{4}{|c|}{Final conventional beam transport elements} \\
\hline
& Magnetic field gradient [T/m] & Current [A]& \parbox[c]{3cm} {\centering --}\\
\hline
Electromagnetic quadrupole 1 & \parbox[c]{3cm} {\centering 2.0 -- 3.5} & \parbox[c]{3cm} {\centering 34--40} & \parbox[c]{3cm} {\centering --}\\
\hline
Beam steerer 1 & \parbox[c]{3cm} {\centering --} & from -1 to 2 & \parbox[c]{3cm} {\centering --}\\
\hline
Beam steerer 2 & \parbox[c]{3cm} {\centering --} & from -1 to 0 & \parbox[c]{3cm} {\centering --} \\
\hline
Electromagnetic quadrupole 2 & \parbox[c]{3cm} {\centering 2.0 -- 3.5} & \parbox[c]{3cm} {\centering 40--46} & \parbox[c]{3cm} {\centering --}\\
\hline
\end{tabular}
\end{table*}


\subsection{The dosimetric systems} \label{dosimetricSystem}

\begin{figure*}
    \centering
    \includegraphics[width=0.8\textwidth]{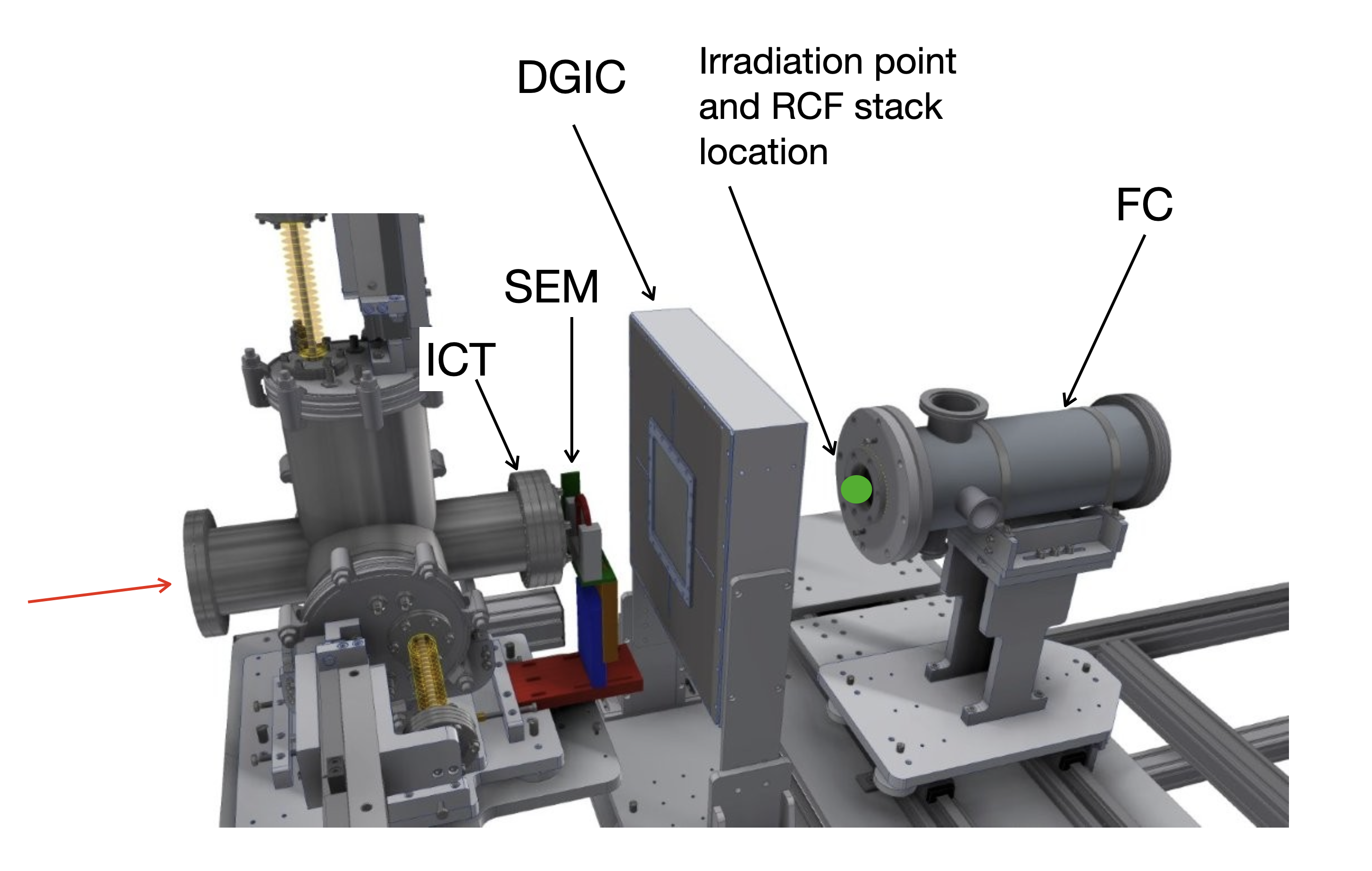}
    \caption{Three-dimensional drawing of the in-air ELIMED beamline section with the dosimetric elements. The particle beam travels from the left (red arrow) and sequentially crosses the ICT, the SEM, and the DGIC (blue dashed box) before reaching the irradiation point, located at the entrance window of the Faraday Cup (green spot).}
    \label{fig:absolute_realtive_setup}
\end{figure*}

The ELIMED dosimetric system was designed to measure the absorbed dose to water at a defined irradiation point and comprised four complementary detectors operating along the in-air section. Figure~\ref{fig:absolute_realtive_setup} illustrates their spatial arrangement, while Table~\ref{tab:device_characteristics} summarizes their functions, positions, operating environments, and signal types. Positions in Table~\ref{tab:device_characteristics} are referenced to the Kapton window ($0~\mathrm{cm}$).

\begin{table*}[ht]
\centering
\caption{Summary of dosimetric devices and their characteristics. Position is measured from the Kapton window.}
\label{tab:device_characteristics}
\begin{tabular}{|l|l|l|l|l|}
\hline
\textbf{Device} & \textbf{Function} & \textbf{Position [cm]} & \textbf{Environment} & \textbf{Output Type} \\
\hline
ICT  & Beam monitoring     & 0   & Vacuum & Total fluence     \\
SEM  & Beam monitoring        & 3.5 & Air    & Relative fluence  \\
DGIC & Calibrated beam monitoring    & 10 & Air    & Ionization current \\
FC   & Absolute dosimeter    & 20 & Air    & Absolute fluence / dose     \\
RCF  & Relative dosimeter    & 20 & Air    & Optical density   \\
\hline
\end{tabular}
\end{table*}

The ICT, manufactured by Bergoz Instrumentation (model ICT-CF6''-60.4.40-UHV-070-5.0-LD)~\cite{ict}, was installed in vacuum immediately upstream of the Kapton window to measure the proton fluence at the air-extraction point, acting as a non-destructive diagnostic that does not perturb the beam. The SEM~\cite{cirrone2020elimed}, consisting of a $7~\mu\mathrm{m}$-thick electrically insulated tantalum foil, was positioned in air $3.5~\mathrm{cm}$ downstream of the window. In addition to providing a relative beam-fluence measurement, it acts as a controlled scatterer that broadens the transverse dose distribution at the irradiation point. Immediately downstream of the SEM, at $10~\mathrm{cm}$ from the window, the DGIC~\cite{detector}, composed of two adjacent free-air ionization chambers with inter-electrode spacings of $5~\mathrm{mm}$ and $10~\mathrm{mm}$, was installed. In the following, these chambers are referred to as DGIC-1 ($5~\mathrm{mm}$) and DGIC-2 ($10~\mathrm{mm}$). Once calibrated in terms of dose to water at the irradiation point, DGIC-1 and DGIC-2 served as online dose monitors; their design enables shot-by-shot recombination correction, ensuring response independence from instantaneous dose rate~\cite{giordanengo2022fluence}. Finally, the FC~\cite{leanza2017faraday} was positioned with its entrance window at $20~\mathrm{cm}$ from the Kapton window, coincident with the irradiation point, to measure the absolute proton fluence on a per-shot basis. In addition, stacks of EBT3 radiochromic films, previously calibrated in terms of dose to water using conventional $30~\mathrm{MeV}$ proton beams~\cite{cirrone2020use}, were employed to evaluate the beam spot size and reconstruct proton energy spectra. Knowledge of the proton fluence, energy, and effective spot size enabled the absolute determination of absorbed dose to water at the irradiation point on a shot-by-shot basis; RCFs also provided an independent verification of the delivered dose.

To characterize the energy distribution of the transported proton beam, a diamond detector operated in time-of-flight (TOF) mode was positioned downstream of the ELIMED magnetic chicane, immediately upstream of the Kapton exit window. The detector was based on a $500~\mu\mathrm{m}$ active diamond layer and was operated with an applied bias voltage of $400~\mathrm{V}$. The use of diamond detectors for TOF diagnostics of high-energy laser-driven ion beams has been described in detail by Scuderi et al.~\cite{scuderi2017time}.

\subsection{Signal processing and absolute-dose determination at the irradiation point}
\label{sec:signal_process}

For each laser shot, the output of each detector was reduced to an integrated quantity: charge for the ICT, DGIC, and FC, and optical density for the RCF. The FC served as the absolute fluence reference detector. The integrated charge $Q$ (C) was converted into absorbed dose to water, $D_w$ (Gy), using the relation:

\begin{equation}
D_w=\frac{S(E)_w}{A}\frac{Q}{e}\cdot 1.602\times10^{-10} \ [Gy]
\label{eq:DoseCalculationFromFC}
\end{equation}

where $A$ (cm$^2$) is the effective beam area, $S(E)_w$ (MeV$\cdot$cm$^2\cdot$g$^{-1}$) is the mean mass stopping power of protons in water at the irradiation point, and $e$ is the elementary charge~\cite{petringaabsolute}. 

For the FC dose conversion, the effective transverse area was calculated from the RCF-derived field size at the 50 $\%$ isodose level. In the present analysis
$A = \pi (d_{\rm eq}/2)^2$, with $d_{\rm eq}=5.5~\mathrm{mm}$.
The same geometrical definition must be kept consistent with the FC acceptance radius adopted in the G4ELIMED simulations (see Section~\ref{sec:DoseVerification}).

The relative uncertainty associated with the FC-derived absorbed dose was evaluated by propagating the main quantities entering Eq.~(\ref{eq:DoseCalculationFromFC}), according to
\begin{equation}
\left(\frac{u(D_w)}{D_w}\right)^2 =
\left(\frac{u(Q)}{Q}\right)^2 +
\left(\frac{u(S_w)}{S_w}\right)^2 +
\left(\frac{u(A)}{A}\right)^2 .
\label{eq:FC_uncertainty}
\end{equation}
Here, $u(Q)$, $u(S_w)$, and $u(A)$ are the standard uncertainties associated with the FC charge measurement, the proton mass stopping power in water, and the effective transverse beam area, respectively.

The uncertainty on the effective beam area was derived from the uncertainty on the beam diameter measured from the radiochromic-film dose distribution (Section~\ref{sec:LateralDoseDistributions}). Assuming a circular irradiation field, $A=\pi(d/2)^2$, the corresponding relative uncertainty is
\begin{equation}
\frac{u(A)}{A}=2\frac{u(d)}{d}.
\label{eq:area_uncertainty}
\end{equation}
Using an effective beam diameter of approximately $5.5~\mathrm{mm}$ and an uncertainty of $0.3~\mathrm{mm}$, this term gives a relative contribution of about $10.9\%$.

The stopping-power contribution was estimated from the reconstructed proton energy spectrum. The spectrum was centred at $23.45~\mathrm{MeV}$ with a FWHM of $2.60~\mathrm{MeV}$ (see Section~\ref{sec:DepthDoseProfile}), corresponding to an equivalent standard deviation of approximately $1.10~\mathrm{MeV}$ if a Gaussian energy distribution is assumed. The resulting uncertainty on $S(E)_w$ was estimated from the local variation of the proton mass stopping power in water around the mean energy and contributes approximately $4\%$ to the FC dose uncertainty. Combining the beam-area and stopping-power terms in quadrature gives a relative uncertainty of approximately $12\%$ on the FC-derived absorbed dose. Any additional independent uncertainty associated with the FC charge readout can be added in quadrature through the first term of Eq.~(\ref{eq:FC_uncertainty}).

The DGIC was cross-calibrated against the FC dose measurements to provide real-time dose evaluation, with recombination corrected on a shot-by-shot basis~\cite{giordanengo2022fluence}. RCFs were analysed using the calibration coefficients and workflow reported in~\cite{cirrone2020use}. 

The electronic front ends of both the ICT and the SEM operate in peak-detection mode. In this mode, the current generated by the detector during the very short proton bunch, of the order of a few nanoseconds, is integrated by the amplifier, which rapidly reaches a maximum value before slowly decreasing during the discharge phase. The peak amplitude is proportional to the total integrated charge and can be converted into coulombs. For the ICT, this provides a measurement of the total charge contained in each proton bunch. For the SEM, it provides a relative indication of the proton fluence.

\begin{figure}
    \centering
    \includegraphics[width=\linewidth]{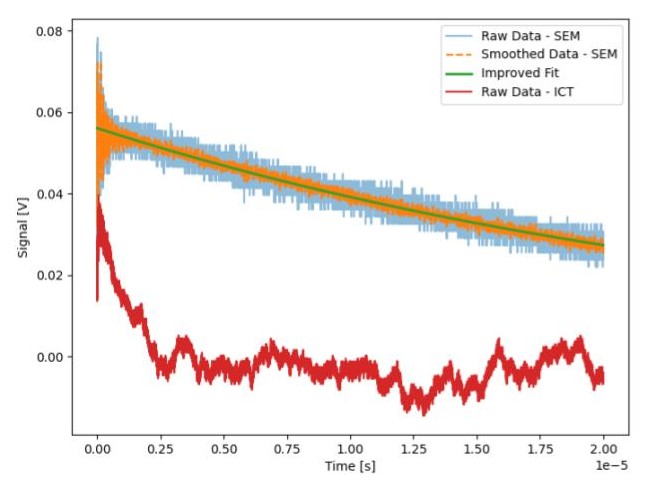}
    \caption{Typical oscilloscope signals recorded from the ICT (red line) and SEM (blue line) after the arrival of a proton bunch. The maximum signal amplitude is converted into the integrated charge measured by each detector.}
    \label{fig:signal_processing2}
\end{figure}

Figure~\ref{fig:signal_processing2} shows the ICT (red curve) and SEM signals acquired by the oscilloscope after the arrival of a laser-driven proton bunch. In both cases, the initial peak is clearly detectable. However, the SEM signal, which is more affected by electronic noise and pulse distortion, was smoothed and fitted in the peak region to determine the pulse height accurately. The smoothed SEM signal and the corresponding fit are also shown in Figure~\ref{fig:signal_processing2}.

\subsection{Irradiation and analysis workflow}
\label{sec:Workflow}

The irradiation campaign followed a fixed workflow designed to provide a traceable measurement of the absorbed dose to water at the user irradiation point (Figure~\ref{fig:absolute_realtive_setup}). 

Beam alignment and basic characterization at the irradiation point were first verified using calibrated EBT3 radiochromic films arranged in a stack configuration. The films were positioned in front of the FC Kapton window, downstream of the plastic collimator, enabling reconstruction of the transverse dose profiles, evaluation of lateral beam symmetry, and measurement of the Bragg curve at the irradiation point. The same EBT3 stack was also used to determine the proton energy spectrum, which is required for the correct estimation of the dose derived from the FC measurements (Section~\ref{sec:signal_process} and Eq.~\ref{eq:DoseCalculationFromFC}).

For each laser shot, the signals from the ICT, SEM, DGIC, and FC were then acquired simultaneously. The integrated charge measured by the FC was converted into absorbed dose to water using Eq.~\ref{eq:DoseCalculationFromFC}, based on the effective spot area estimated from the radiochromic-film measurements.
The DGIC signals were subsequently cross-calibrated against the FC on a shot-by-shot basis, applying recombination corrections to obtain an online absolute dose monitor (Sec.~\ref{sec:absolute}). Finally, the ICT and SEM waveforms, recorded in peak-detection mode, were used as relative fluence monitors and correlated with the FC and DGIC measurements (Sec.~\ref{sec:absolute}).

    \section{Results}
    \label{sec:results}
    
\subsection{Relative dosimetry at the irradiation point}
\label{sec:relative_dosimeter}

Before the absolute dosimetric calibration at the irradiation point, a relative dosimetric characterization of the proton beam was performed. This step was required to determine the transverse beam distribution and the depth--dose profile, from which the main beam parameters entering the absolute dose evaluation could be derived. In particular, the transverse dose distribution was used to estimate the effective beam spot area at the irradiation point, which is required for evaluating the dose from the FC charge. The depth--dose profile and reconstructed energy spectrum were used to determine the mean proton energy and the corresponding mass stopping power in water, thereby reducing the uncertainty associated with the FC-derived absorbed dose.
    
\subsubsection{Lateral dose distributions}
\label{sec:LateralDoseDistributions}

At the irradiation point (see Figure~\ref{fig:absolute_realtive_setup}), transverse dose profiles were measured using a stack of calibrated EBT3 radiochromic films exposed to 100 consecutive laser-driven proton bunches.

\begin{figure}
    \centering
    \includegraphics[width=\linewidth]{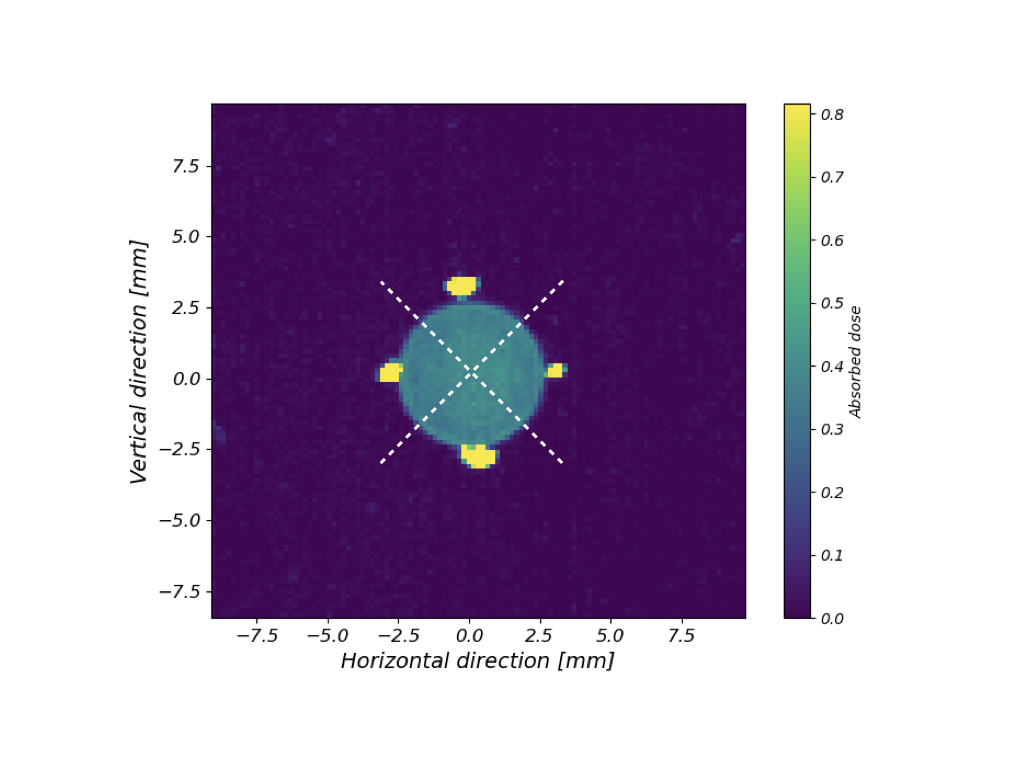}
    \caption{Two-dimensional absorbed-dose distribution measured on the first EBT3 radiochromic film layer of the stack positioned at the irradiation point, downstream of a 5~mm diameter circular collimator. The colour map represents the absorbed dose to water, expressed in gray (Gy), delivered by 100 consecutive laser-driven proton bunches. The yellow features visible on the film are artefacts due to pen marks made during the alignment procedure to indicate the physical position and aperture of the collimator, and should not be interpreted as part of the dose distribution. The two orthogonal dashed white lines indicate the regions from which the lateral dose profiles shown in Figure~\ref{fig:trasversalDoseProfiles} were extracted.}
    \label{fig:beamSpotProfile}
\end{figure}

Figure~\ref{fig:beamSpotProfile} shows the two-dimensional beam-spot profile recorded on the first radiochromic film of the stack. The colour scale represents the absorbed dose to water, expressed in gray (Gy), with dark blue corresponding to the lowest dose values. The pixel intensities were converted into absorbed dose to water using a previously established calibration procedure~\cite{cirrone2020use}.

Two orthogonal lateral dose profiles were extracted from the distribution shown in Figure~\ref{fig:beamSpotProfile} along the two dashed white lines. Both profiles were normalized to their respective maximum values and centred at the zero coordinate, corresponding to the geometrical axis of the proton beam.

\begin{figure}
    \centering
    \includegraphics[width=\linewidth]{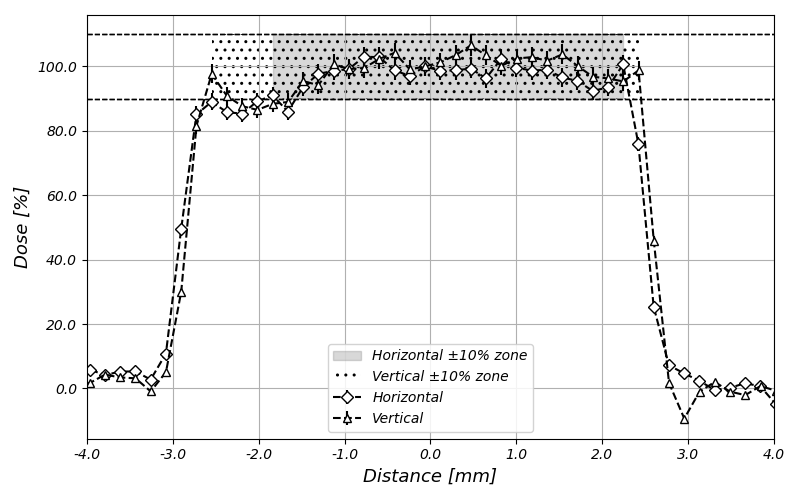}
    \caption{Normalized lateral dose profiles extracted along the two orthogonal transverse directions indicated in Figure~\ref{fig:beamSpotProfile}. The lateral penumbra is defined as the distance over which the dose decreases from 80\% to 20\% of the maximum value. The field size corresponds to the lateral extent of the beam at the 50\% isodose level. The horizontal bars indicate the regions of dose homogeneity within a $\pm10\%$ tolerance around the central axis for each profile.}
    \label{fig:trasversalDoseProfiles}
\end{figure}

\begin{table}[h]
\centering
\begin{tabular}{lcc}
\hline
\textbf{Parameter} & \textbf{Horizontal} & \textbf{Vertical} \\
\hline
\makecell[c]{Left penumbra (80--20\%) \\ {[mm]}}   & 0.29 & 0.24 \\
\makecell[c]{Right penumbra (80--20\%) \\ {[mm]}}  & 0.35 & 0.18 \\
\makecell[c]{Field size (50\%--50\%) \\ {[mm]}}    & $5.50 \pm 0.30$ & $5.43 \pm 0.30$ \\
\makecell[c]{Homogeneity at 10\% \\ {[mm]}}        & $4.08 \pm 0.30$ & $4.97 \pm 0.30$ \\
\makecell[c]{Homogeneity at 5\% \\ {[mm]}}         & $2.04 \pm 0.30$ & $2.48 \pm 0.30$ \\
\hline
\end{tabular}
\caption{Quantitative parameters extracted from the lateral dose distributions shown in Figure~\ref{fig:trasversalDoseProfiles}.}
\label{tab:lateral_distribution_parameters}
\end{table}

The transverse dose distributions were characterized in terms of lateral penumbra, field size, and dose homogeneity, as summarized in Table~\ref{tab:lateral_distribution_parameters}. The beam exhibits a sharp lateral dose fall-off, with 80--20\% penumbra values below 0.4~mm in both transverse directions. Minor asymmetries are observed between the left and right penumbrae and between the horizontal and vertical profiles; however, all values remain within the sub-millimetric range, indicating effective beam collimation and limited lateral scattering.

The field size, defined at the 50\% isodose level, is approximately $5.5 \pm 0.3~\mathrm{mm}$ in both transverse directions. The relative difference between the horizontal and vertical field sizes is below 2\%, confirming the overall geometrical symmetry of the irradiation field.

Dose homogeneity within the central region of the field was evaluated using 10\% and 5\% tolerance criteria. Homogeneous regions of $4.08 \pm 0.30~\mathrm{mm}$ and $4.97 \pm 0.30~\mathrm{mm}$ were obtained in the horizontal and vertical directions, respectively, within a 10\% dose variation. When applying the more stringent 5\% homogeneity criterion, the usable field size is reduced to approximately 2.0--2.5~mm. These results indicate that the beam provides a sufficiently uniform dose distribution over a millimetric region, suitable for controlled \textit{in vitro} irradiation experiments.

Overall, the combination of sharp lateral penumbrae, well-defined field size, and adequate central dose uniformity confirms the suitability of the laser-driven proton beam for applications requiring precise lateral dose confinement.
    
\subsubsection{Depth--dose profile and energy spectrum}
\label{sec:DepthDoseProfile}

After characterization of the lateral dose distribution, the same RCF stack positioned at the irradiation point was used to reconstruct the depth--dose profile in water and infer the energy spectrum of the incident protons. The dose deposited along the stack provided high-spatial-resolution information on the longitudinal beam properties. This reconstruction was based on the previously established film calibration and on a dedicated analysis approach developed to retrieve both depth--dose information and proton energy spectra from the experimental data~\cite{cirrone2020elimed,guarrera2024proton}.

\begin{figure}
    \centering
    \includegraphics[width=0.95\linewidth]{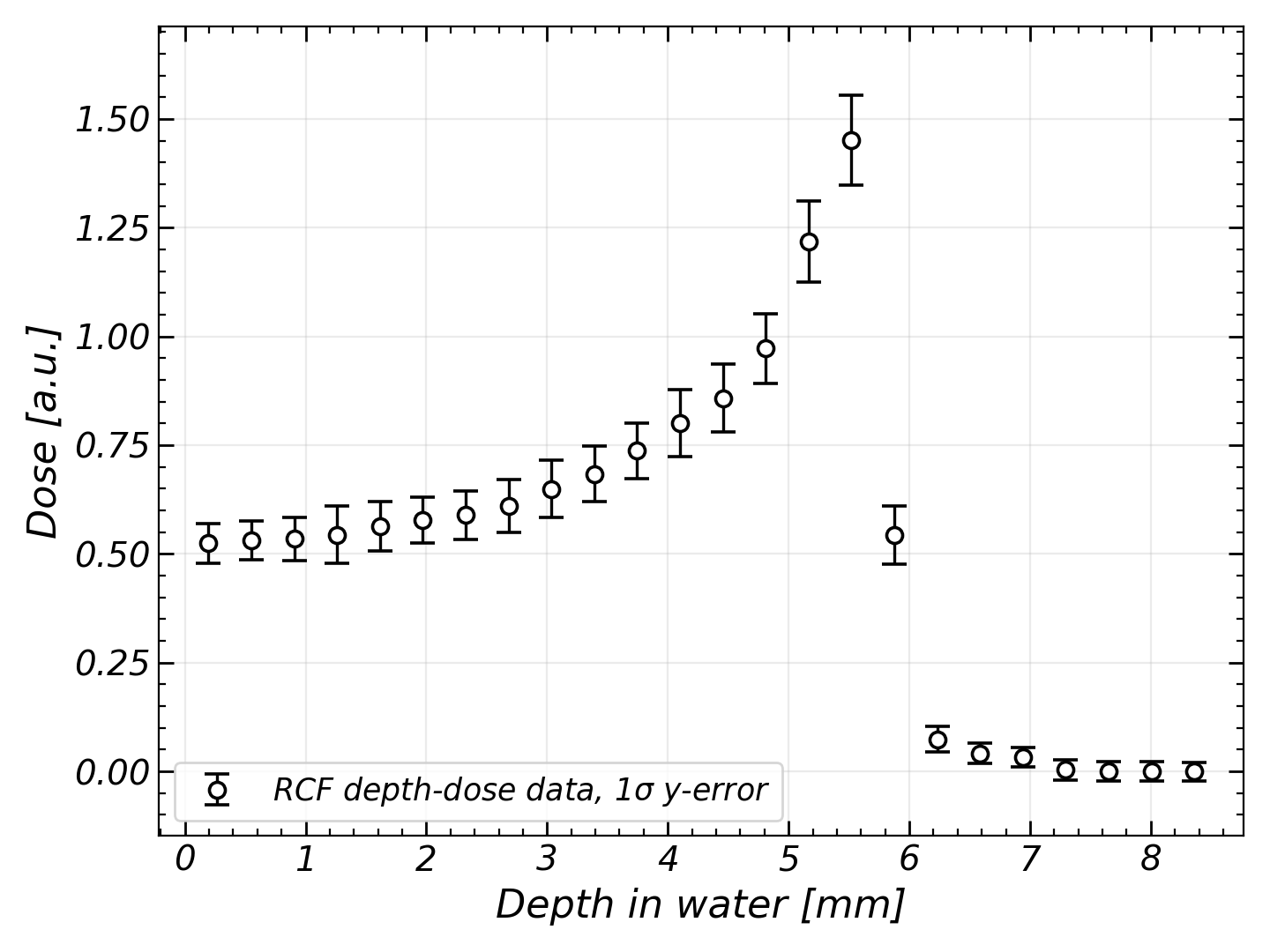}
    \caption{Reconstructed depth--dose distribution in water obtained from the radiochromic film stack positioned at the irradiation point. The profile shows the characteristic proton beam behaviour, with an entrance plateau followed by a gradual dose increase and a Bragg peak located at a water-equivalent depth of approximately $5.5$--$5.7~\mathrm{mm}$, corresponding to an incident mean proton energy of $23.45 \pm 0.5~\mathrm{MeV}$. The finite width of the Bragg peak reflects the intrinsic energy spread of the energy-selected laser-driven proton beam, as well as energy-loss straggling within the film stack.}
    \label{fig:depth_dose_distribution}
\end{figure}

The depth--dose distribution reconstructed from the radiochromic film stack and normalized to the entrance dose is shown in Figure~\ref{fig:depth_dose_distribution}. The profile exhibits the characteristic behaviour of a proton beam propagating in water, with an initial plateau followed by a gradual dose increase with depth, reflecting the increase in stopping power as the proton energy decreases. A Bragg peak is clearly visible at a water-equivalent depth of approximately $5.5$--$5.7~\mathrm{mm}$.

The mean incident proton energy was estimated from the practical range, defined as the depth corresponding to the distal dose fall-off at 10\% of the peak value~\cite{IAEA2021_TRS398}. In the present case, a practical range of $6.3 \pm 0.3~\mathrm{mm}$ in water was found, corresponding to an incident proton energy of $23.45 \pm 0.5~\mathrm{MeV}$. The main characteristics of the measured depth--dose distribution are reported in Table~\ref{tab:depth_dose_distribution}.

\begin{table}[h!]
\centering
\caption{Summary of the main depth--dose distribution parameters.}
\label{tab:depth_dose_distribution}
\begin{tabular}{l l}
\hline
\textbf{Parameter} & \textbf{Value} \\
\hline
Bragg peak position & $5.52 \pm 0.30~\mathrm{mm}$ \\
Peak-to-plateau ratio & 2.8 \\
Full width at half maximum & $2.1 \pm 0.3~\mathrm{mm}$ \\
Distal fall-off (80\%--20\%) & $0.19 \pm 0.30~\mathrm{mm}$ \\
Practical range (R$_{10}$) & $6.3 \pm 0.3~\mathrm{mm}$ \\
\hline
\end{tabular}
\end{table}

Beyond the Bragg peak, the dose decreases rapidly. However, the Bragg peak exhibits a finite width that is broader than expected for a highly monoenergetic proton beam. This broadening is mainly attributed to the intrinsic energy spread introduced by the energy selection system and, in particular, by the finite acceptance of the energy-selector slits. Additional contributions may arise from energy-loss straggling within the RCF stack.

Overall, the measured depth--dose profile is consistent with expectations for an energy-selected, non-monoenergetic laser-driven proton beam. These results confirm the capability of the radiochromic film stack to reconstruct longitudinal dose distributions and to provide quantitative information on the proton energy spread under realistic experimental conditions.

\begin{figure}
    \centering
    \includegraphics[width=1\linewidth]{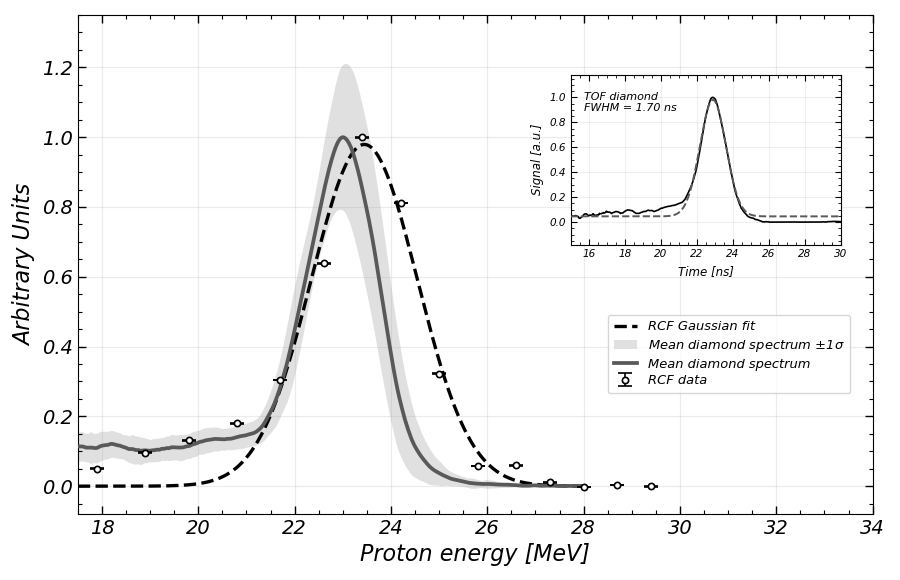}
    \caption{
Normalized proton energy spectra compared at the RCF irradiation plane. The RCF-reconstructed spectrum is shown together with a Gaussian fit to the main peak, from which a mean energy of $23.45~\mathrm{MeV}$ and an energy FWHM of $2.60~\mathrm{MeV}$ were obtained. Vertical error bars represent the $1\sigma$ uncertainties on the reconstructed RCF spectrum. The mean diamond-TOF spectrum was measured in vacuum at the diamond detector position and is shown after energy-loss propagation to the RCF plane, accounting for the Kapton exit window, the SEM tantalum foil, and the air path between the two measurement planes. The grey band represents the shot-to-shot standard deviation of the selected diamond-TOF spectra. After this propagation, the residual separation between the two peak positions is about $480~\mathrm{keV}$, corresponding to approximately $2\%$. The inset shows the corresponding time-domain diamond-TOF signal in the proton-bunch region; the dashed curve represents a Gaussian fit, yielding a temporal FWHM of $1.70~\mathrm{ns}$. The agreement between the two independent spectral measurements supports the consistency of the RCF-based energy reconstruction.
}

    \label{fig:spectrum}
\end{figure}

Using the same irradiated radiochromic film stack and a deconvolution procedure adapted from Kaufman et al.~\cite{kaufman2015radiochromic}, the proton energy spectrum at the irradiation point was reconstructed, as shown in Figure~\ref{fig:spectrum}. The spectrum is centred at approximately $23.45~\mathrm{MeV}$, in good agreement with the mean proton energy inferred from the practical range of the depth--dose distribution~\cite{cirrone2020elimed,guarrera2024proton}. The main peak was fitted with a Gaussian function, yielding a full width at half maximum (FWHM) of $2.60~\mathrm{MeV}$. To provide an independent check of the spectral reconstruction and, at the same time, access the temporal structure of the transported proton bunch, the RCF spectrum was compared with diamond time-of-flight (TOF) measurements acquired during the same commissioning configuration. The diamond detector was located in vacuum, just upstream of the air-extraction window, whereas the RCF stack measured the spectrum at the irradiation point in air. Therefore, the two diagnostics refer to different longitudinal planes along the beamline. The diamond-TOF spectra were averaged over the selected shots and normalized to their maximum value; their shot-to-shot dispersion is represented by the standard-deviation band in Figure~\ref{fig:spectrum}. For the spectral comparison, the TOF-derived energy distribution was propagated from the diamond plane to the RCF plane through an energy-loss calculation that accounts for proton energy loss in the Kapton exit window, the SEM tantalum foil, and the intervening air path. After this propagation, the two peak positions differ by about $480~\mathrm{keV}$, corresponding to approximately $2\%$.

The comparison shown in Figure~\ref{fig:spectrum} provides a cross-check of the proton spectrum obtained with the RCF stack. The RCF measurement gives the energy spectrum at the irradiation point through an integrating dosimetric detector, whereas the diamond-TOF diagnostic provides an independent online measurement of the spectral distribution at an upstream in-vacuum plane and, through its time-domain signal, of the bunch temporal profile. Since the two spectra refer to different measurement planes, the expected energy loss between the diamond detector and the RCF stack was included through the energy-loss propagation described above. The residual peak separation of about $480~\mathrm{keV}$ is small compared with the mean beam energy and supports the consistency of the RCF-reconstructed spectrum used for the subsequent dose determination.

The temporal width extracted from the diamond-TOF signal also provides a route to estimate the instantaneous peak dose rate of a single laser-driven proton bunch. The diamond detector does not provide an absolute dose measurement by itself; however, assuming that the temporal dose profile is proportional to the measured diamond-TOF signal and can be approximated by a Gaussian function, the peak dose rate can be estimated by combining the temporal FWHM with the independently measured dose per pulse. For a Gaussian temporal profile,
\begin{equation}
\dot{D}_{\mathrm{peak}} = \frac{D_{\mathrm{pulse}}}{\sigma_t\sqrt{2\pi}},
\label{eq:peak_dose_rate_sigma}
\end{equation}
where $D_{\mathrm{pulse}}$ is the dose delivered by a single proton bunch and $\sigma_t$ is the temporal standard deviation of the bunch. Since
\begin{equation}
\sigma_t = \frac{\mathrm{FWHM}_t}{2\sqrt{2\ln2}},
\label{eq:sigma_time_fwhm}
\end{equation}
Eq.~(\ref{eq:peak_dose_rate_sigma}) can be written as
\begin{equation}
\dot{D}_{\mathrm{peak}} \simeq 0.94\,\frac{D_{\mathrm{pulse}}}{\mathrm{FWHM}_t}.
\label{eq:peak_dose_rate_fwhm}
\end{equation}
Using the RCF dose measured in the independent fifty-shot irradiation, $D_{\mathrm{RCF}}=36.46~\mathrm{cGy}$, the corresponding dose per pulse is $D_{\mathrm{pulse}}=7.29\times10^{-3}~\mathrm{Gy}$. With the temporal width extracted from the diamond-TOF inset, $\mathrm{FWHM}_t=1.70~\mathrm{ns}$, Eq.~(\ref{eq:peak_dose_rate_fwhm}) gives
\begin{equation}
\dot{D}_{\mathrm{peak}} \simeq 4.0\times10^{6}~\mathrm{Gy\,s^{-1}}.
\label{eq:peak_dose_rate_value}
\end{equation}
This value should be regarded as an estimate of the peak dose rate within an individual proton bunch under the present commissioning conditions; the average dose rate instead depends on the number of bunches delivered and on the laser repetition rate.

From the reconstructed spectrum, a total proton fluence of approximately $6.18 \times 10^{7}~\mathrm{cm}^{-2}$ was derived, corresponding to an average fluence of about $6.2 \times 10^{5}~\mathrm{cm}^{-2}$ per proton bunch within the quasi-uniform beam spot shown in Figure~\ref{fig:beamSpotProfile}.

Taken together, the lateral dose uniformity reported in Table~\ref{tab:lateral_distribution_parameters}, the measured depth--dose characteristics summarized in Table~\ref{tab:depth_dose_distribution}, and the reconstructed energy spectrum provide a coherent assessment of the beam quality at the irradiation point. These results validate the use of the present experimental configuration for the subsequent absolute-dose measurements and radiobiological irradiation studies.    
\subsection{Absolute dosimetry and beam monitoring of the dose-delivery system}
\label{sec:absolute}

The absorbed dose to water measured by the FC at the irradiation point was adopted as the absolute reference dosimetric quantity. This reference dose, evaluated on a shot-by-shot basis using the signal-processing workflow described in Sec.~\ref{sec:Workflow}, was used to cross-calibrate the DGIC installed along the in-air section of the beamline, as shown in Figure~\ref{fig:absolute_realtive_setup}. In this configuration, the FC and the DGIC constitute the primary dosimetric chain of the dose-delivery system: the FC provides the absolute dose reference at the irradiation point, while the DGIC provides the online in-air monitor cross-calibrated against it.

\begin{figure*}
    \centering
    \includegraphics[width=\linewidth]{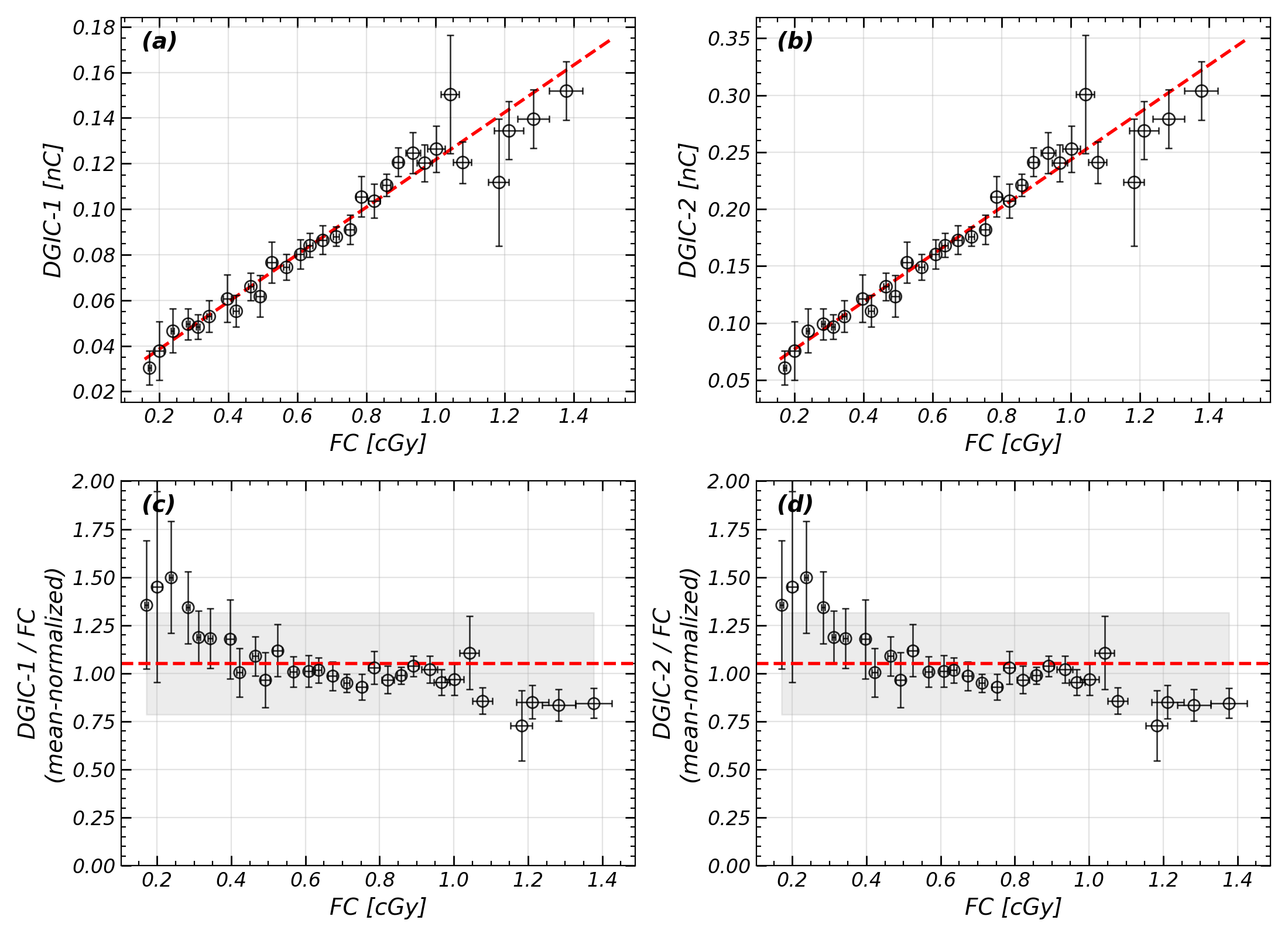}
    
    \caption{
Cross-calibration of the two DGIC gaps against the absorbed dose measured by the Faraday Cup at the irradiation point. 
For graphical clarity, the experimental data are grouped into FC-dose bins; each plotted point represents the mean value of the shots contained in the corresponding bin, while the error bars represent the associated binned uncertainties, including the propagated uncertainty on the FC-derived dose. 
(a) Recombination-corrected charge collected by DGIC-1 as a function of the FC dose. 
(b) Recombination-corrected charge collected by DGIC-2 as a function of the FC dose. 
In panels (a) and (b), the dashed lines represent linear fits performed on the full unbinned dataset. 
(c) Mean-normalized DGIC-1-to-FC response ratio as a function of the FC dose. 
(d) Mean-normalized DGIC-2-to-FC response ratio as a function of the FC dose. 
In panels (c) and (d), the horizontal dashed line indicates the mean response, while the shaded band represents the corresponding $\pm1\sigma$ interval. 
Both DGIC gaps show an approximately linear response over the investigated dose range, with DGIC-2 exhibiting about twice the sensitivity of DGIC-1.
}
    \label{fig:DGIC_FC}
\end{figure*}

For the graphical representation of the DGIC--FC cross-calibration, the experimental points were grouped as a function of the FC dose in order to reduce visual overlap while preserving the statistical behaviour of the full dataset. The valid shots were divided into 40 equally spaced dose intervals, and only bins containing at least two shots were displayed. Each point shown in Figure~\ref{fig:DGIC_FC} therefore represents the mean value of all measurements falling within the corresponding dose bin. The horizontal error bars represent the uncertainty associated with the binned FC dose, while the vertical error bars represent the uncertainty associated with the binned DGIC charge.

The horizontal uncertainty bars associated with the FC dose include the propagated uncertainty on the absolute dose determination according to Eq.~(\ref{eq:FC_uncertainty}). In the present analysis, the dominant contribution arises from the uncertainty on the effective beam area, while the finite energy spread contributes through the corresponding variation of the proton stopping power in water. The resulting relative uncertainty on the FC-derived absorbed dose was estimated to be approximately $12\%$.

For each DGIC gap, an effective absolute uncertainty on the collected charge was estimated from the standard deviation of the residuals obtained from the linear fit to the full unbinned dataset. For each dose bin, the uncertainty assigned to the displayed mean value was conservatively taken as the larger of the standard error of the mean of the points within the bin and the propagated measurement uncertainty. The linear fits, however, were performed on the full unbinned dataset and not on the binned points, so that the extracted sensitivities remain representative of the complete set of valid shots.

Figure~\ref{fig:DGIC_FC} summarizes the correlation between the integrated charge collected in each DGIC gap and the corresponding absorbed dose measured by the FC at the irradiation point, together with the associated mean-normalized DGIC-to-FC response ratios. Over the explored dose range, both gaps, DGIC-1 and DGIC-2, exhibit an approximately linear response to the reference FC dose.

The dose-binned representation shown in Figure~\ref{fig:DGIC_FC} preserves the overall trend of the complete shot-by-shot dataset while reducing the visual crowding caused by the large number of individual laser shots. Panels (a) and (b) show that both DGIC gaps exhibit an approximately linear response with respect to the FC-derived absorbed dose. Linear regression performed on the full unbinned dataset yields sensitivities of $0.1039~\mathrm{nC/cGy}$ for DGIC-1 and $0.2077~\mathrm{nC/cGy}$ for DGIC-2, with coefficients of determination of $R^{2}=0.731$ for both gaps. These values are summarized in Table~\ref{tab:sensitivity}. The factor-of-two difference between the two sensitivities is consistent with the doubled collection gap of DGIC-2 with respect to DGIC-1 under comparable operating conditions.

Panels (c) and (d) report the mean-normalized DGIC-to-FC response ratios, defined as the ratio between the DGIC signal normalized to its mean value and the FC dose normalized to its mean value. The uncertainty bars in these panels were obtained by propagating the uncertainties associated with the DGIC signal and the FC-derived dose, while treating the normalization factors as fixed scaling constants. The larger relative uncertainties observed at the lowest dose values reflect the reduced signal-to-noise ratio in this region: since the DGIC uncertainty includes an approximately constant absolute component estimated from the residual dispersion around the fit, its relative contribution increases as the collected charge decreases. Despite this behaviour, the binned ratios remain distributed around unity without evidence of a systematic dose-dependent trend, supporting the use of both DGIC gaps as online dose monitors for the ELIMAIA--ELIMED beamline.

\begin{table}[h!]
    \centering
    \begin{tabular}{lcc}
    \hline
    \textbf{Device} & \textbf{Sensitivity} & \textbf{$R^2$ coefficient} \\
    & \textbf{[nC/cGy]} & \\
    \hline
    DGIC-1 & 0.1039 & 0.731 \\
    DGIC-2 & 0.2077 & 0.731 \\
    \hline
    \end{tabular}
    \caption{Sensitivity of the two DGIC gaps obtained from the linear regression against the FC-derived absorbed dose.}
    \label{tab:sensitivity}
\end{table}

After establishing the FC--DGIC dosimetric chain, the SEM and ICT signals were analysed as upstream beam-monitoring quantities. In contrast to the DGIC, these detectors were not used as primary online dose monitors in the present calibration procedure. Instead, their response was evaluated to assess their capability to track relative shot-to-shot variations in proton-beam intensity and to provide consistency checks along the beamline.

\begin{figure}[ht!]
    \centering
    \includegraphics[width=1\linewidth]{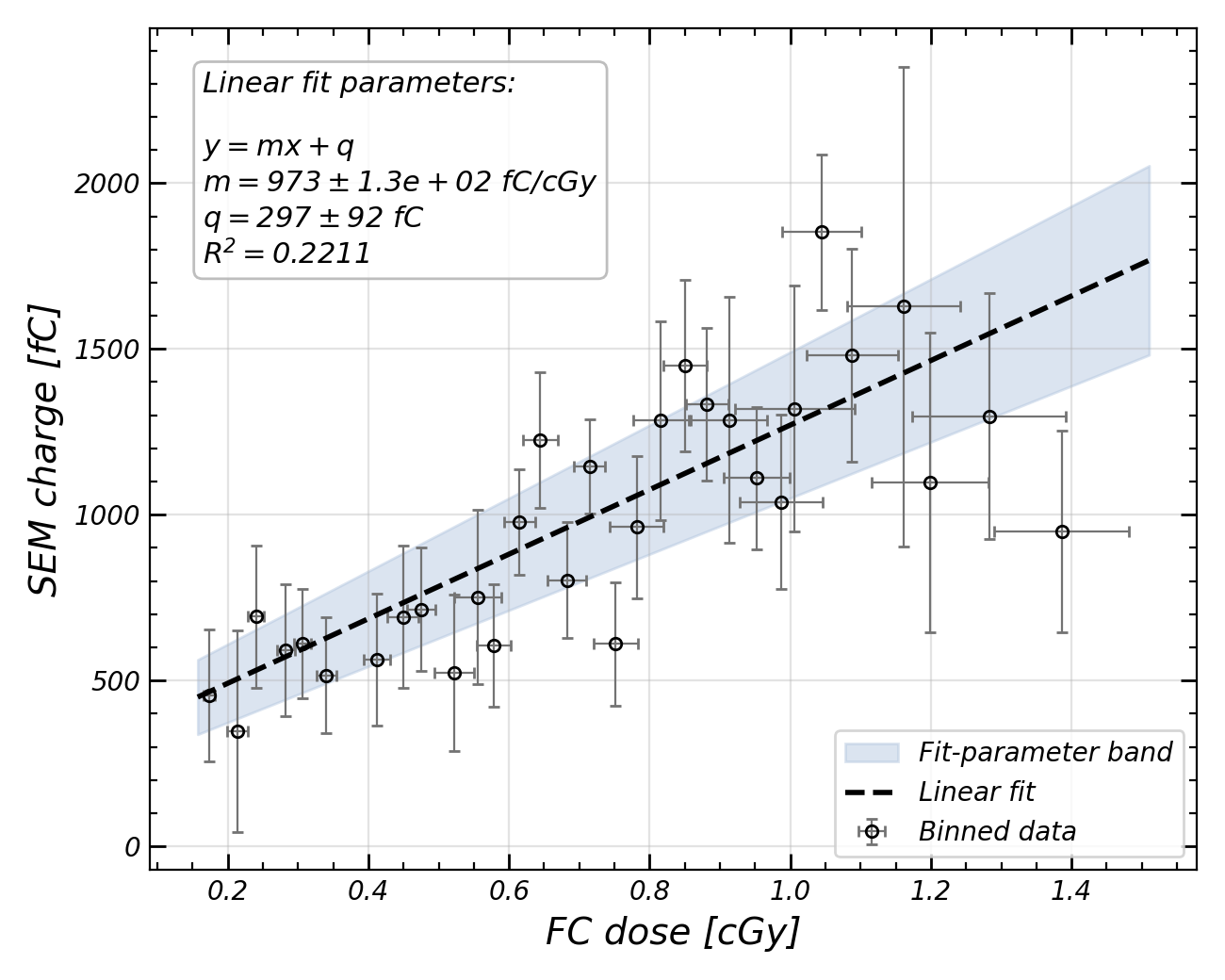}\\
    \caption{
Correlation between the charge collected by the Secondary Electron Monitor (SEM) and the absorbed dose to water derived from the Faraday Cup (FC) at the irradiation point. 
The experimental data were grouped into FC-dose bins for graphical clarity, with each point representing the mean FC dose and the mean SEM charge within the corresponding bin. 
Horizontal error bars include the propagated uncertainty on the FC-derived dose, while vertical error bars represent the uncertainty on the binned SEM charge. 
The dashed line represents the linear fit performed on the full selected unbinned dataset, while the shaded band indicates the uncertainty envelope associated with the fit parameters.
}
    \label{fig:SEM-vs-FC}
\end{figure}

\begin{figure}[ht!]
    \centering
    \includegraphics[width=1\linewidth]{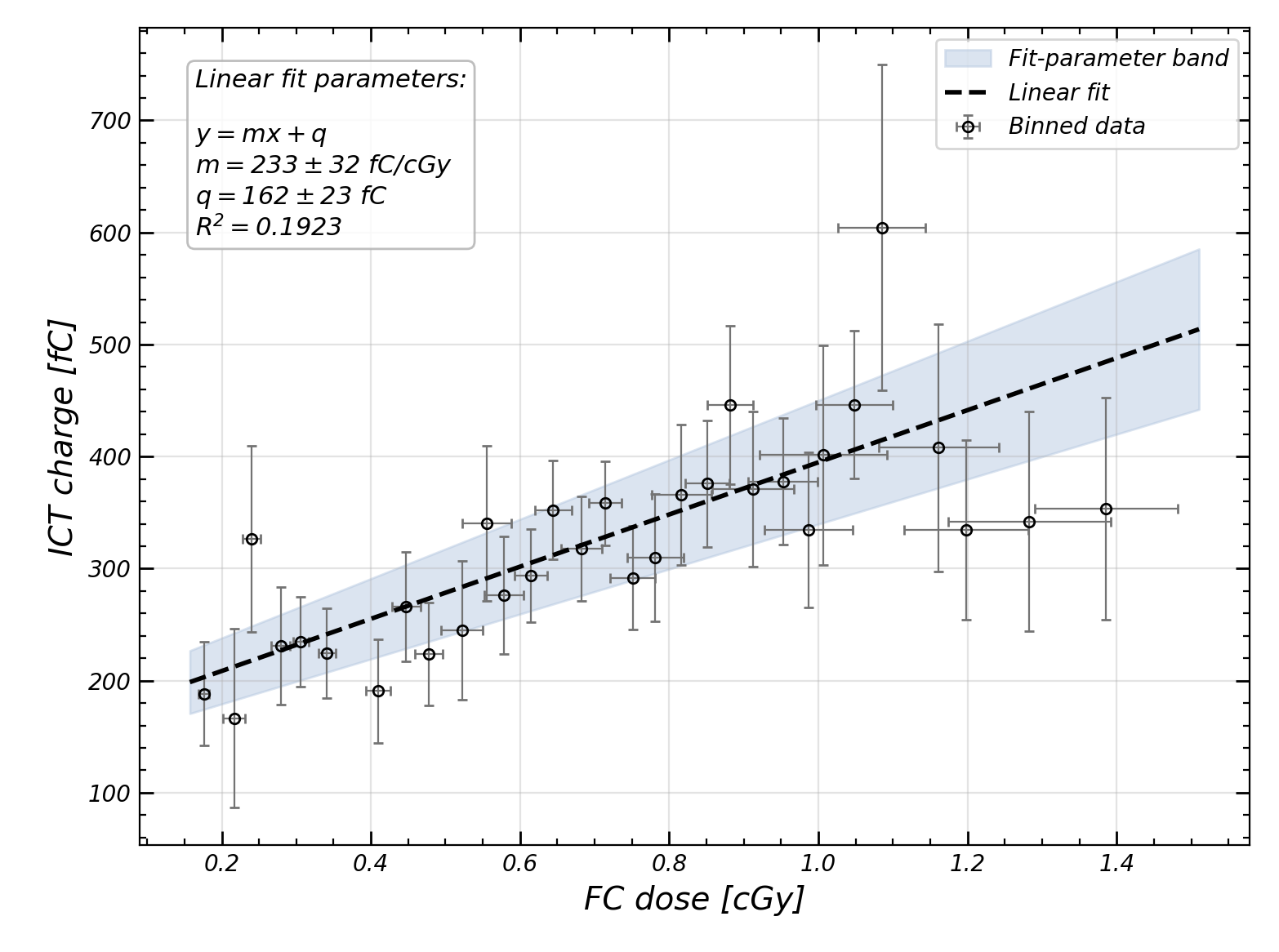}\\
    \caption{
Correlation between the charge collected by the Integrating Current Transformer (ICT) and the absorbed dose to water derived from the Faraday Cup (FC) at the irradiation point. 
The experimental data were grouped into FC-dose bins for graphical clarity, with each point representing the mean FC dose and the mean ICT charge within the corresponding bin. 
Horizontal error bars include the propagated uncertainty on the FC-derived dose, while vertical error bars represent the uncertainty on the binned ICT charge. 
The dashed line represents the linear fit performed on the full selected unbinned dataset, while the shaded band indicates the uncertainty envelope associated with the fit parameters.
}
    \label{fig:ICT-vs-FC}
\end{figure}

The performance of the SEM and ICT as relative beam monitors was first evaluated by correlating their shot-by-shot responses with the absorbed dose measured by the FC at the irradiation point, as shown in Figure~\ref{fig:SEM-vs-FC} for the SEM and in Figure~\ref{fig:ICT-vs-FC} for the ICT. In both cases, the horizontal error bars represent the propagated uncertainty on the FC-derived absorbed dose, estimated to be approximately $12\%$. For graphical clarity, the data were grouped into FC-dose bins. Each plotted point represents the mean FC dose and the mean detector charge of the shots contained in the corresponding bin, while the fit was performed on the full selected unbinned dataset.

The SEM--FC correlation shown in Figure~\ref{fig:SEM-vs-FC} is weak, with a coefficient of determination of approximately $R^2=0.221$. Similarly, the ICT--FC correlation shown in Figure~\ref{fig:ICT-vs-FC} exhibits a weak linear correlation, with $R^2=0.192$. These results indicate that, under the low-fluence conditions investigated during the present commissioning campaign, neither the SEM nor the ICT provides a sufficiently robust standalone estimate of the dose delivered at the irradiation point on a shot-by-shot basis. This behaviour is mainly attributed to the small collected signals, for which electronic noise, statistical fluctuations, and possible beam-transport variations between the monitor positions and the FC can represent a significant fraction of the measured response.

\begin{figure}[ht!]
    \centering
    \includegraphics[width=1\linewidth]{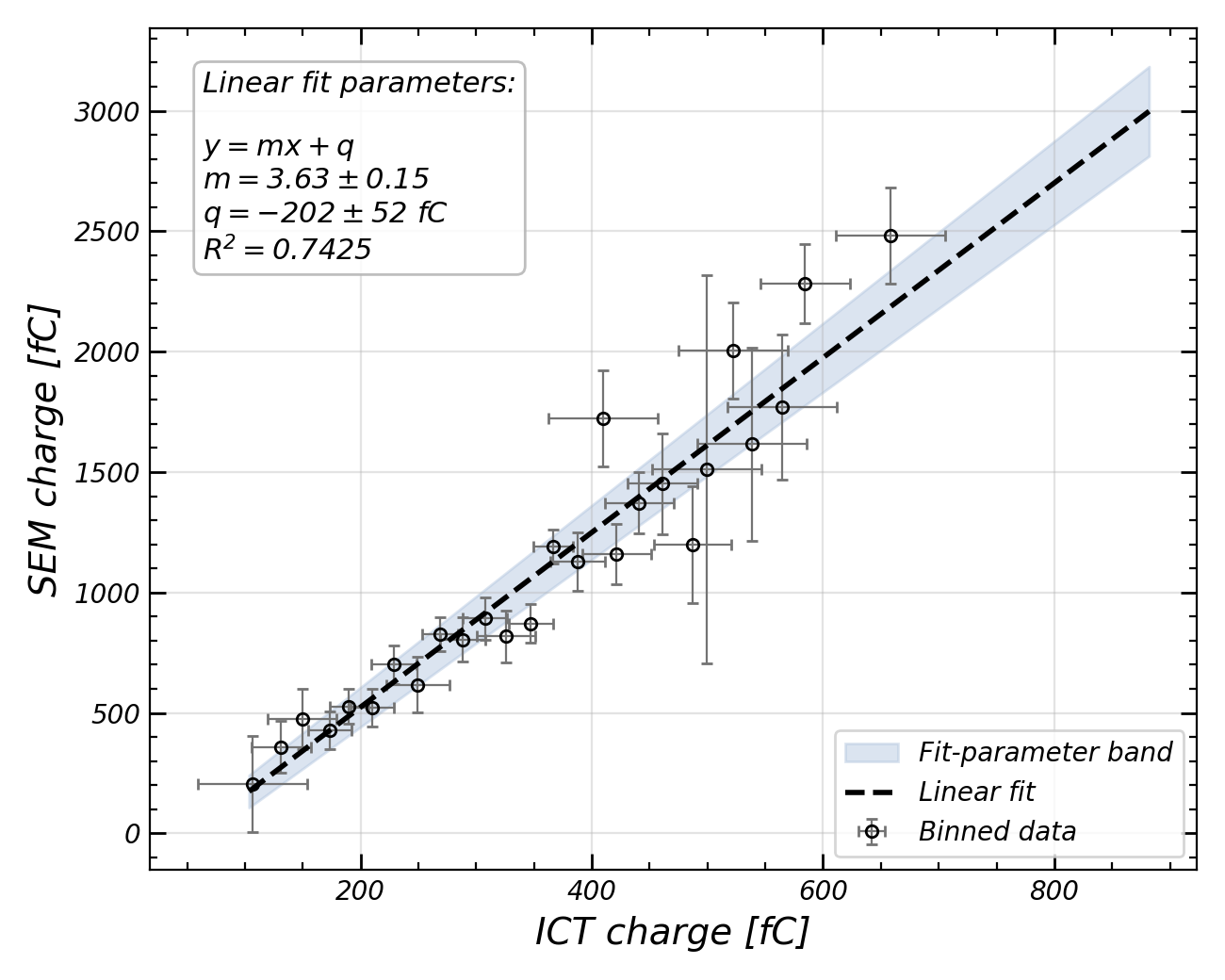}\\
    \includegraphics[width=1\linewidth]{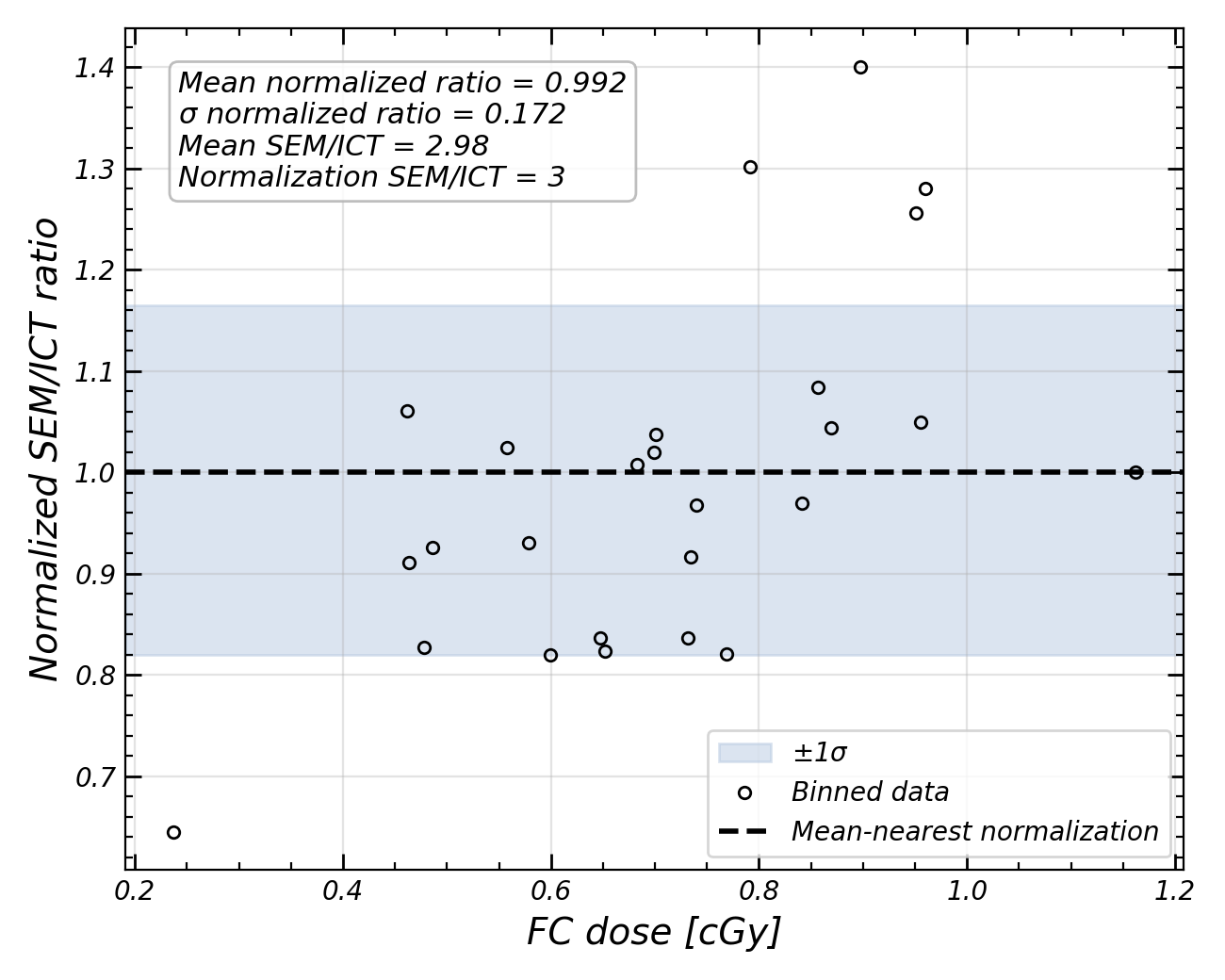}
    \caption{
Correlation between the signals measured by the Integrating Current Transformer (ICT) and the Secondary Electron Monitor (SEM). 
Top panel: SEM charge as a function of ICT charge. The data are grouped into ICT-charge bins for graphical clarity, with each point representing the mean ICT and SEM signals within the corresponding bin. The dashed black line represents the linear fit performed on the full selected unbinned dataset, while the shaded band indicates the uncertainty envelope associated with the fit parameters. 
Bottom panel: normalized SEM/ICT ratio as a function of the FC-derived dose. The ratio was calculated using the same binned points shown in the top panel and was normalized to the point whose SEM/ICT value is closest to the mean ratio. The dashed horizontal line indicates the normalized reference value, while the shaded band represents the $\pm 1\sigma$ interval of the normalized ratio distribution.
}
    \label{fig:ICT_SEM}
\end{figure}

The direct correlation between the ICT and SEM responses recorded for the same set of laser shots was then analysed, as shown in Figure~\ref{fig:ICT_SEM}-top. In this case, a clearer monotonic and approximately linear relationship is observed, with a coefficient of determination of approximately $R^2=0.742$. This indicates that, although the individual correlations of the SEM and ICT with the FC-derived dose are weak, the two upstream monitors are mutually consistent in tracking relative variations in the proton-beam intensity.

Figure~\ref{fig:ICT_SEM}-bottom shows the normalized SEM/ICT signal ratio as a function of the FC-derived dose. The ratio was calculated using the same binned points shown in Figure~\ref{fig:ICT_SEM}-top and was normalized to the point whose SEM/ICT value is closest to the mean ratio. Within the experimental dispersion, the normalized ratio remains distributed around unity without evidence of a clear systematic dose-dependent trend. This result supports the use of the SEM and ICT as relative beam-monitoring devices and as mutual consistency checks, despite their limited capability to provide an independent quantitative dose estimate under the present low-fluence conditions.

Operationally, the direct ICT--SEM correlation can be useful for monitoring shot-to-shot beam-intensity variations, identifying anomalous shots, and providing redundancy against transient readout instabilities. However, in the present configuration, the absolute dose delivery remains anchored to the FC--DGIC dosimetric chain. The SEM and ICT should therefore be regarded as complementary upstream beam monitors rather than as primary dosimetric devices.

Taken together, the pairwise correlations in Figures~\ref{fig:DGIC_FC}, \ref{fig:SEM-vs-FC}, \ref{fig:ICT-vs-FC}, and~\ref{fig:ICT_SEM} establish a coherent hierarchy for the ELIMAIA--ELIMED dose-delivery system: the FC provides the absolute reference at the irradiation point, the DGIC provides the primary online dose monitor cross-calibrated against the FC, and the SEM and ICT provide upstream relative beam-monitoring information.

\subsection{Independent verification of the delivered dose at the irradiation point}
\label{sec:DoseVerification}
\label{single}

For independent verification of the absolute dose determination, a dedicated irradiation consisting of fifty consecutive laser shots was performed. A single calibrated EBT3 radiochromic film was positioned at the irradiation point, directly upstream of the FC, while the FC remained operational. This configuration allowed the absorbed dose to water to be determined simultaneously from the charge collected by the FC and from the response of the EBT3 film.

Using the measured proton energy spectrum and the effective beam spot area, the total charge $Q$ collected by the FC was converted into absorbed dose to water according to Eq.~(\ref{eq:DoseCalculationFromFC}). The resulting cumulative dose delivered over the fifty shots was
\[
D_{\mathrm{FC}} = 30.45 \pm 3.5~\mathrm{cGy},
\]
where the uncertainty includes the contributions from the effective beam area and from the stopping-power variation associated with the measured proton energy spread.

The corresponding dose measured using the EBT3 film was
\[
D_{\mathrm{RCF}} = 36.46 \pm 1.8~\mathrm{cGy}.
\]
The uncertainty on the RCF dose includes contributions from the film calibration, scanner response, film uniformity, and dose-conversion procedure. Considering the central values, the dose derived from the FC is approximately $16.5\%$ lower than the value independently obtained with the EBT3 film, when the difference is evaluated with respect to the RCF dose.

To investigate the origin of this discrepancy, dedicated Monte Carlo simulations were performed with the G4ELIMED~\cite{G4ELIMED_1, G4ELIMED_2} application, based on the Geant4 toolkit~\cite{Geant42006,Geant42016}. The simulations reproduced the RCF--FC comparison geometry used during the fifty-shot irradiation, including the EBT3 film positioned immediately upstream of the FC, the air gap between the film and the FC entrance, and the geometrical acceptance of the FC. The proton source was defined according to the measured energy spectrum at the irradiation point, with mean energy $23.45~\mathrm{MeV}$ and FWHM $2.60~\mathrm{MeV}$, and according to the measured transverse field size reported in Table~\ref{tab:lateral_distribution_parameters}. The same source was transported in two configurations: without the RCF, corresponding to propagation through the air gap only, and with the RCF inserted upstream of the FC. A total of $6\times10^{6}$ primary proton histories was simulated for the reference configuration.

The simulation scored the transverse proton distribution at the FC entrance plane and the fraction of protons entering the FC geometrical acceptance. These two scored quantities were used to estimate, respectively, the RCF-induced increase of the effective beam area and the relative transmission loss due to protons scattered outside the FC acceptance.

The relative transmission was evaluated as
\[
T_{\mathrm{rel}} =
\frac{T_{\mathrm{with\,RCF}}}{T_{\mathrm{without\,RCF}}},
\]
where $T_{\mathrm{with\,RCF}}$ and $T_{\mathrm{without\,RCF}}$ are the fractions of protons entering the FC acceptance with and without the RCF, respectively. The simulation gave
\[
T_{\mathrm{rel}} \simeq 0.945,
\]
corresponding to a relative proton loss of approximately $5.5\%$ due to scattering induced by the RCF. The corresponding correction for transmission losses is therefore
\[
C_{T} = \frac{1}{T_{\mathrm{rel}}} \simeq 1.058.
\]

The second effect is the increase in the transverse beam size at the FC entrance. The angular broadening induced by multiple Coulomb scattering in the RCF redistributes part of the proton fluence towards the lateral regions of the beam, producing a larger effective transverse spread at the FC entrance than in the case without the film. Since the FC-derived absorbed dose depends on the effective beam area entering Eq.~(\ref{eq:DoseCalculationFromFC}), this broadening must be taken into account when comparing the FC-derived dose with the RCF dose measured at the irradiation point.

In this work, the RMS transverse beam sizes are defined as the standard deviations of the simulated proton spatial distributions along the two transverse coordinates at the FC entrance plane. Specifically, $\sigma_x$ and $\sigma_y$ are calculated from the full distributions of the proton positions in the horizontal and vertical directions, respectively. These quantities provide a measure of the overall transverse spread of the beam, including the contribution of particles scattered into the lateral tails, and are therefore more appropriate than the 50\% profile width for estimating the effective beam-area increase induced by the RCF.

The corresponding RMS effective transverse area was evaluated as
\[
A_{\mathrm{eff}} = 4\pi\sigma_x\sigma_y ,
\]
where $\sigma_x$ and $\sigma_y$ are the RMS transverse beam sizes. For a uniform circular beam, this definition reduces to the geometrical beam area; for the present simulated distributions, it provides a consistent estimate of the relative area increase associated with RCF-induced multiple Coulomb scattering. The area correction can therefore be written as
\[
C_{A} =
\frac{A_{\mathrm{with\,RCF}}}{A_{\mathrm{without\,RCF}}}
=
\frac{\sigma_{x,\mathrm{with\,RCF}}\sigma_{y,\mathrm{with\,RCF}}}
{\sigma_{x,\mathrm{without\,RCF}}\sigma_{y,\mathrm{without\,RCF}}}.
\]
The calculation gave
\[
C_{A} \simeq 1.048,
\]
corresponding to an increase of the effective transverse area of approximately $4.8\%$.

\begin{figure}[t]
\centering
\includegraphics[width=0.82\linewidth]{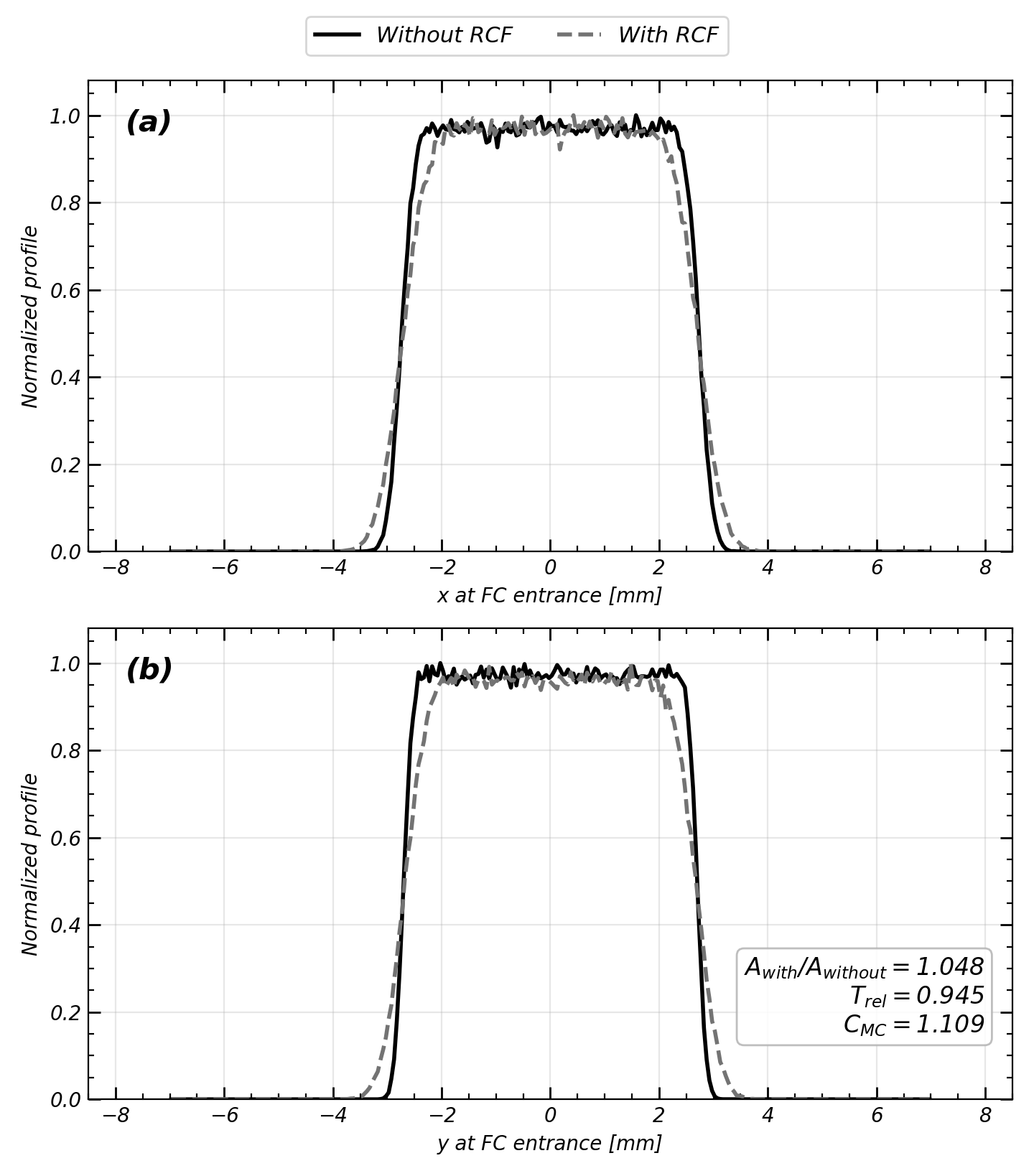}
\caption{
Monte Carlo evaluation of the transverse beam redistribution induced by the EBT3 film positioned upstream of the FC. Panel (a) shows the horizontal central-slice profile at the FC entrance, while panel (b) shows the corresponding vertical profile. The black solid curves refer to the transport without the RCF, whereas the grey dashed curves include the RCF followed by the air gap to the FC entrance. Each profile is independently normalized to its own maximum value in order to highlight the RCF-induced modification of the beam shape. The profiles are shown for visualization only: the effective-area correction was evaluated from the RMS transverse beam sizes of the full simulated distributions, while the absolute proton loss outside the FC geometrical acceptance was obtained from the simulated relative transmission.
}
\label{fig:MC_beam_broadening}
\end{figure}

Figure~\ref{fig:MC_beam_broadening} shows the simulated transverse central-slice profiles at the FC entrance. The profiles were independently normalized to their respective maximum values; consequently, the figure is intended to visualize the RCF-induced redistribution of the beam shape, not the absolute proton loss. The grey dashed profiles show the edge smearing and the formation of lateral tails caused by multiple Coulomb scattering in the RCF. Since the curves are independently normalized and represent central-slice profiles, their 50\% widths are not used to determine the area correction. The effective-area correction was obtained quantitatively from the RMS transverse beam sizes of the full simulated distributions, while the transmission correction was obtained independently by counting the fraction of protons entering the FC geometrical acceptance.

The overall Monte Carlo correction factor was obtained by combining the transmission correction and the effective-area correction,
\[
C_{\mathrm{MC}} =
\frac{C_A}{T_{\mathrm{rel}}}
=
\frac{1.048}{0.945}
\simeq 1.109.
\]
This factor accounts for both the proton losses outside the FC acceptance and the increase of the effective beam area at the FC entrance caused by scattering in the RCF. Applying this correction factor to the FC measurement gives
\[
D_{\mathrm{FC,corr}} = C_{\mathrm{MC}} D_{\mathrm{FC}}
                    = 33.80 \pm 3.90~\mathrm{cGy},
\]
where the quoted uncertainty was obtained by propagating only the uncertainty on $D_{\mathrm{FC}}$. 

The main quantities entering the RCF--FC comparison are summarized in Table~\ref{tab:rcf_fc_verification}. Before applying the Monte Carlo correction, the FC-derived dose is lower than the RCF dose by approximately $16.5\%$. After applying the correction factor, the residual difference is reduced to approximately $7.3\%$, evaluated with respect to the RCF dose.

\begin{table*}[t]
\centering
\small
\renewcommand{\arraystretch}{1.15}
\begin{tabular}{p{0.28\textwidth} p{0.24\textwidth} p{0.40\textwidth}}
\hline
\textbf{Quantity} & \textbf{Value} & \textbf{Comment} \\
\hline
FC-derived dose 
& $30.45 \pm 3.5~\mathrm{cGy}$ 
& Uncorrected FC value. \\

RCF dose 
& $36.46 \pm 1.8~\mathrm{cGy}$ 
& Independent EBT3 measurement. \\

Difference FC--RCF 
& $16.5\%$ 
& Difference between the uncorrected FC dose and the RCF dose, evaluated with respect to the RCF dose. \\

Relative FC transmission 
& $T_{\mathrm{rel}} \simeq 0.945$ 
& Relative fraction of protons entering the FC acceptance with the RCF compared with the case without the RCF. \\

Transmission correction 
& $C_T = 1/T_{\mathrm{rel}} \simeq 1.058$ 
& Correction associated with protons scattered by the RCF outside the FC geometrical acceptance. \\

Effective-area correction 
& $C_A \simeq 1.048$ 
& Correction associated with the RCF-induced increase of the RMS effective transverse beam area at the FC entrance. \\

Total Monte Carlo correction factor 
& $C_{\mathrm{MC}} \simeq 1.109$ 
& Combined correction accounting for both transmission losses and beam-area enlargement. \\

Corrected FC dose 
& $33.8 \pm 3.9~\mathrm{cGy}$ 
& FC-derived dose after application of the Monte Carlo correction factor. \\

Residual difference 
& $\sim 7.3\%$ 
& Residual difference between the corrected FC dose and the RCF dose, evaluated with respect to the RCF dose. \\
\hline
\end{tabular}
\caption{
Summary of the independent RCF--FC dose verification performed at the irradiation point. The FC-derived dose was corrected using a Monte Carlo correction factor accounting for the loss of protons outside the FC geometrical acceptance and for the increase of the RMS effective transverse beam area at the FC entrance due to multiple Coulomb scattering in the RCF. The uncertainty reported for the corrected FC dose includes only the propagated uncertainty on the uncorrected FC dose; The uncertainty reported for the corrected FC dose includes the propagated uncertainty on the uncorrected FC dose only; the uncertainty associated with the Monte Carlo correction factor was not included in this estimate.
}
\label{tab:rcf_fc_verification}
\end{table*}

These results indicate that the observed discrepancy between the FC and RCF dose determinations is largely consistent with the perturbation introduced by the RCF itself when positioned immediately upstream of the FC. In particular, multiple Coulomb scattering in the film causes both a partial loss of protons outside the FC geometrical acceptance and a broadening of the beam spot at the FC entrance. After accounting for these effects, the FC- and RCF-based dose estimates agree within approximately $7\%$ in terms of central values.

From an operational point of view, the commissioning identifies a clear hierarchy of the dosimetric system under the present low-fluence conditions. The FC provides the traceable absolute reference at the irradiation point, the DGIC provides the primary online dose monitor once cross-calibrated against the FC, and the ICT and SEM provide upstream relative beam-monitoring information. The residual dispersion observed in the detector correlations is mainly associated with the low collected charge and shot-to-shot fluctuations of the laser-driven source. Future higher-fluence operation is therefore expected to reduce the relative impact of electronic noise and to improve the precision of the online monitor calibration.
\section{Conclusion}\label{conclusion}

This work presents the first relative and absolute dosimetric commissioning of the ELIMAIA--ELIMED laser-driven proton irradiation beamline at ELI Beamlines. The commissioning campaign established and validated the complete dosimetric chain from the beam extraction point to the user irradiation station, combining absolute and relative dosimetry through a set of complementary detectors including a Faraday Cup (FC), Dual-Gap Ionization Chamber (DGIC), Integrating Current Transformer (ICT), Secondary Electron Monitor (SEM), and calibrated radiochromic films (RCFs).

The transported proton beam was characterized in terms of spatial dose distribution, depth--dose profile, energy spectrum, fluence, delivered absorbed dose, and bunch temporal structure. At the irradiation point, a symmetric irradiation field of approximately $5.5~\mathrm{mm}$ diameter was obtained, with sub-millimetric lateral penumbrae and a homogeneous central region suitable for small-field radiobiological applications. The depth--dose measurements and spectrum reconstruction performed with RCF stacks yielded an average proton energy of $23.4~\mathrm{MeV}$ with a FWHM of approximately $2.6~\mathrm{MeV}$, in agreement with the proton range derived from the measured Bragg curve. An independent diamond-TOF measurement acquired upstream in vacuum provided a complementary spectral check. After energy-loss propagation from the diamond plane to the RCF irradiation plane, accounting for the Kapton exit window, the SEM tantalum foil, and the air path, the TOF-derived and RCF-reconstructed peak positions differed by about $480~\mathrm{keV}$, corresponding to approximately $2\%$. The same diamond-TOF signal provided a temporal FWHM of $1.70~\mathrm{ns}$ for the selected proton bunch. Combining this temporal width with the measured dose per pulse yielded an estimated peak dose rate of approximately $4.0\times10^{6}~\mathrm{Gy\,s^{-1}}$ under the present commissioning conditions.

Absolute dose measurements were performed using the FC as the reference detector and were used to cross-calibrate the DGIC for online dose monitoring. Both DGIC chambers exhibited a linear response with respect to the FC-derived dose over the investigated range, with sensitivities of $0.1039$ and $0.2077~\mathrm{nC/cGy}$ for DGIC-1 and DGIC-2, respectively. The ICT and SEM also showed a consistent mutual correlation, confirming their suitability as upstream relative beam-fluence monitors and providing a redundant monitoring system along the beamline.

The absolute dose determination was independently verified using calibrated EBT3 radiochromic films. The uncorrected FC dose was lower than the RCF dose by approximately $16.5\%$. Dedicated Monte Carlo simulations performed with the G4ELIMED application showed that the presence of the RCF upstream of the FC introduces both a relative transmission loss, $T_{\mathrm{rel}}\simeq0.945$, and an increase of the effective transverse beam area at the FC entrance, $C_A\simeq1.048$. The resulting correction factor, $C_{\mathrm{MC}}\simeq1.109$, reduced the residual difference between the corrected FC dose and the RCF dose to approximately $7\%$ in terms of central values. Considering the quoted FC and RCF uncertainties, the corrected values are compatible within the present experimental uncertainty.

Overall, the results demonstrate that the ELIMAIA--ELIMED beamline is capable of delivering laser-driven proton beams with controlled energy selection, well-characterized spatial and spectral properties, and traceable absolute dosimetry at the user irradiation point. The successful commissioning of the monitoring and dosimetric systems establishes the experimental framework required for forthcoming radiobiological investigations and represents an important milestone toward the routine use of laser-driven proton beams for biomedical and medical-physics applications. Preliminary radiobiological irradiations were carried out shortly after this first commissioning campaign~\cite{Blaha2025}.

It should be noted that the present commissioning was performed under conditions of relatively low proton fluence per shot at the irradiation point. Consequently, several detectors operated close to their lower sensitivity limits, and shot-to-shot statistical fluctuations contributed significantly to the observed dispersion in the detector correlations and to the overall experimental uncertainty. Despite these challenging conditions, the complementary dosimetric systems exhibited consistent behaviour and enabled a complete characterization of the transported beam and delivered dose. This outcome confirms the robustness of the adopted dosimetric methodology under the present commissioning conditions.

From an operational point of view, the commissioning identifies a clear hierarchy of the dosimetric system. The FC provides the traceable absolute reference at the irradiation point, the DGIC provides the primary online dose monitor once cross-calibrated against the FC, and the ICT and SEM provide upstream relative beam-monitoring information. This detector hierarchy is essential for future user experiments, where online monitoring, shot-to-shot control, and independent absolute dose verification will have to operate together as a single irradiation and dosimetric system.

Future experimental campaigns will be carried out with substantially higher proton fluence and improved source stability, enabling a significant reduction of statistical uncertainties and a more accurate assessment of detector performance. These measurements will allow a more precise determination of detector sensitivities, linearity, repeatability, and uncertainty budgets, as well as the extension of the commissioning to different beam energies, irradiation configurations, and ultra-high-dose-rate conditions.

A central objective of future developments will be to demonstrate controlled multi-pulse irradiation schemes at the user sample, with dose delivery conditions approaching those required for fast-fractionation radiobiology. In this perspective, operation at higher repetition rate, with doses per pulse of the order of $0.1~\mathrm{Gy}$, average dose rates approaching $1~\mathrm{Gy\,s^{-1}}$, and instantaneous dose rates of the order of $10^{8}~\mathrm{Gy\,s^{-1}}$ should be considered as target performance to be demonstrated in forthcoming campaigns, rather than as results of the present commissioning.

Achieving these parameters with traceable dosimetry, adequate field uniformity, and reproducible shot-to-shot delivery would enable systematic studies of biological responses as a function of dose per pulse, pulse spacing, accumulated dose, and instantaneous dose rate. Combined with the international user-access model of the ELI research infrastructure, these developments would further consolidate ELIMAIA--ELIMED as a dedicated user platform for quantitative research with laser-driven proton beams and for future biomedical and medical-physics applications.

\section*{Acknowledgement}
This work is based on experiments performed at the ELIMED station, at the ELI  Beamline Facility, supported through beamtime allocated under the 3rd ELI ERIC User Call and registered under Experiment ID ELIUPM3-44.
We acknowledge the Extreme Light Infrastructure (ELI) for providing access to experimental facilities and support. The use of instrumentation, services (including sample preparation), and experimental support was made possible by the ELI Excellence-based User Programme. We would like to thank Petr Szotkowski, Jakub Novak and the L3 team for the assistance during the experiment.
We acknowledge the INFN-LNS technical for the support in the electronic front-end optimisation.
This work was also partially supported by the Committee V of the INFN.


\bibliographystyle{unsrt}
\bibliography{references}
\end{document}